\documentclass[10pt,sigconf,letterpaper]{acmart}
\usepackage[english]{babel}
\usepackage{blindtext}
\usepackage{caption}
\usepackage{subcaption}
\usepackage{amsmath}
\usepackage{stmaryrd}
\usepackage{tcolorbox}
\usepackage{multirow}

\setcopyright{acmcopyright}
\copyrightyear{2020}
\acmYear{2020}

\renewcommand\footnotetextcopyrightpermission[1]{} %

\begin{document}

\title[Understanding the Evolution of Blockchain Ecosystems]{Understanding the Evolution of Blockchain Ecosystems: A Longitudinal Measurement Study of Bitcoin, Ethereum, and EOSIO}

\author{Ningyu He}
\affiliation{%
 \institution{Peking University}
 \city{Beijing}
 \country{China}
}

\author{Weihang Su}
\affiliation{%
 \institution{Beijing University of Posts and Telecommunications}
 \city{Beijing}
 \country{China}
}

\author{Zhou Yu}
\affiliation{%
 \institution{Beijing University of Posts and Telecommunications}
 \city{Beijing}
 \country{China}
}

\author{Xinyu Liu}
\affiliation{%
 \institution{Beijing University of Posts and Telecommunications}
 \city{Beijing}
 \country{China}
}

\author{Fengyi Zhao}
\affiliation{%
 \institution{Beijing University of Posts and Telecommunications}
 \city{Beijing}
 \country{China}
}

\author{Haoyu Wang*}
\affiliation{%
 \institution{Beijing University of Posts and Telecommunications}
 \city{Beijing}
 \country{China}
}

\author{Xiapu Luo}
\affiliation{%
 \institution{The Hong Kong Polytechnic University}
 \city{Hong Kong}
 \country{China}
}

\author{Gareth Tyson}
\affiliation{%
 \institution{Queen Mary University of London}
 \city{London}
 \country{United Kindom}
}

\author{Lei Wu}
\affiliation{%
 \institution{Zhejiang University}
 \city{Zhejiang}
 \country{China}
}

\author{Yao Guo*}
\affiliation{%
 \institution{Peking University}
 \city{Beijing}
 \country{China}
}

\renewcommand{\shortauthors}{Ningyu, et al.}

\begin{abstract}
The continuing expansion of the blockchain ecosystems has attracted much attention from the research community. However, although a large number of research studies have been proposed to understand the diverse characteristics of individual blockchain systems (e.g., Bitcoin or Ethereum), little is known at a comprehensive level on the evolution of blockchain ecosystems at scale, longitudinally, and across multiple blockchains. 
We argue that understanding the dynamics of blockchain ecosystems could provide  unique insights that cannot be achieved through studying a single static snapshot or a single blockchain network alone.
Based on billions of transaction records collected from three representative and popular blockchain systems (Bitcoin, Ethereum and EOSIO) over 10 years, we conduct the first study on the evolution of multiple blockchain ecosystems from different perspectives. Our exploration suggests that, although the overall blockchain ecosystem shows promising growth over the last decade, a number of worrying ``outliers'' exist that have disrupted its evolution. 
\end{abstract}

\maketitle
\pagestyle{plain}

\section{Introduction}

It has been over 12 years since blockchain was invented to serve as the public transaction ledger of Bitcoin~\cite{bitcoin}. Since, blockchain technologies have evolved significantly. This includes hundreds of public blockchain platforms, thousands of cryptocurrencies~\cite{cryptocurrency}, tens of thousands of decentralized applications (DApps) and blockchain-based services~\cite{dapp}, millions of smart contracts~\cite{he2020characterizing}, hundreds of millions of user accounts~\cite{eos-account}, and billions of transactions~\cite{eth-tx}.
These technologies and entities have formed a symbiotic relationship, which we refer to as the \textit{\textbf{blockchain ecosystem}}.

The global blockchain economy is estimated to be worth 1.8 trillion USD at the end of May 2021~\cite{cryptocurrency}.
As a result, its continuing evolution has increased its complexity, which \textit{itself needs considerable effort to understand: particularly its characteristics, operations, trends and security issues}.

Due to this, the blockchain ecosystem has attracted a lot of attention from the research community. A large number of studies have focused on exploring the characteristics of blockchain networks~\cite{huang2020understanding, chen2020traveling, wu2019t}, detecting attacks and vulnerabilities~\cite{chen2018understanding, suevil, ji2020deposafe}, deanonymizing users and tracking money-flow~\cite{ober2013structure, awan2017blockchain, moser2018empirical, wang2020identifying}, and so on.
However, little is known (at a comprehensive level) on the evolution of the overall blockchain ecosystem. Although several prior studies have performed measurement studies on transactions and smart contracts, they characterize activities based on either a static snapshot~\cite{chen2020traveling}, or based on a single platform~\cite{chen2018understanding, huang2020understanding}. 
For example, Chen et al.~\cite{chen2018understanding} and Huang et al.~\cite{huang2020understanding} characterized Ethereum and EOSIO, respectively, based on large-scale transaction analysis.
To the best of our knowledge, \textit{no existing studies have characterized the evolution of blockchain ecosystems comprehensively at scale, longitudinally and across multiple representative platforms.}

\textbf{This Work.}
To fill this gap, we perform a multi-dimensional and large-scale study covering all of the transactions on three of the most representative blockchains: Bitcoin, Ethereum and EOSIO. 
Bitcoin is the first distributed blockchain implementation.
Ethereum is the second largest blockchain system after Bitcoin, and the first second-generation blockchain platform that supports smart contract functionality~\cite{eth-blockchain-2}.
EOSIO claims to be the third-generation blockchain, and adopts Delegate Proof of Stake (DPoS) consensus, which significantly improves throughput and enables new applications~\cite{eos-dapp-prosperity-1, eos-dapp-prosperity-2}.

We first collect the largest ever dataset across these three platforms (\textbf{see \S\ref{sec:study-design}}), with over 35 billion traces in total. 
We use this dataset to provide a high-level characterization and temporal analysis of the three platforms (\textbf{\S\ref{sec:blockchain-evolution}}), including assessing the behavior of money transfers (\textbf{\S\ref{sec:blockchain-evolution:mtg}}), account creation (\textbf{\S\ref{sec:blockchain-evolution:acg}}) and contract invocation (\textbf{\S\ref{sec:blockchain-evolution:cig}}).
Following this, we explore ``outliers'' across the evolution of the three blockchains. We identify these as either misbehavior by attackers or highly popular ``killer'' DApps that trigger major fluctuations in activity (\textbf{\S\ref{sec:abnormal-behaviors}}).
Among many interesting results, the following are the most prominent:
\begin{itemize}
    \item \textit{The overall blockchain ecosystem has shown significant growth over the last decade. Ethereum and EOSIO are on their way to becoming key decentralized systems, featuring powerful and popular DApps.} 
    Bitcoin has shown a steady upward trend, with the number of transactions rising 15K times over 12 years. The number of transactions in Ethereum and EOSIO rose 7.6K times and 2.2K times, respectively.
    However, along with this growth in the number of transactions, the proportion of money transferred has dropped to 16.5\% and 15.1\% for Ethereum and EOSIO till Nov. 2019, while the proportion of smart contract invocation has reached over 80\%.
    Smart contracts and DApps are therefore becoming increasingly popular.
    
    \item \textit{Many activities in the blockchain ecosystem follow the Pareto principle across their evolution, and we are witnessing increasing centralization.}
    For all three types of activities, i.e., money transfer, account creation, and contract invocation, the Pareto principle is followed (across all blockchains). This applies to every month we analyze, indicating that a small group of accounts are increasingly controlling the entire network. 
    
    \item \textit{Interactions among smart contracts are increasingly popular.}
    During the development of Ethereum and EOSIO, the emergence of ``killer'' DApps has led to geater interactions between smart contracts. 
    In other words, contract invocation transactions are more likely to be initiated by another smart contract than by a human.
    However, Ethereum and EOSIO show quite different behaviors regarding to the DApp evolution. For example, in Ethereum, the explosion of Decentralized Finance (DeFi) DApps accounts for over 1/3 of contract invocations since Mar. 2020. In contrast, nearly half of (49.04\%) EOSIO contract invocation transactions were related to gambling DApps.
    
    \item \textit{87 ``outliers'' are identified from our time-series analysis. These have a significant impact that either facilitates or impedes the progress of the blockchain ecosystem}. 48 of these are due to emerging DApps, while 39 of them are introduced by attacks or scams.
\end{itemize}

To the best of our knowledge, this is the first longitudinal and large-scale study across these three-generations of blockchain. Our results motivate the need for further research efforts. Particularly, we argue it is necessary to measure and mitigate the widely unexplored trends and misbehaviors of these blockchain technologies.

\section{Background}
\label{sec:background}

\subsection{Blockchain and its Evolution}
\label{sec:background:blockchain-evolution}

Blockchain has a history spanning 12 years. 
\textit{Bitcoin}, as the first platform that utilized blockchain, is widely called \textit{Blockchain 1.0}. 
Following the growth of cryptocurrencies, users 
started to explore the possibility of decentralized applications (DApps). This led to the development of Ethereum. As a representative system of \textit{Blockchain 2.0}, \textit{Ethereum} is widely adopted by developers. In fact, millions of smart contracts have already been deployed on Ethereum. However, the performance of Ethereum is limited by its consensus protocol (Proof-of-Work, PoW). As a result, in 2018, EOSIO came online, which adopts Delegated Proof-of-Stake (DPoS) consensus. This has been termed \textit{Blockchain 3.0}, improving transaction throughput significantly. To boost performance, Ethereum 2.0 (Phase 0 test) has also adopted Proof-of-Stake (PoS) consensus in Dec. 2020.
Although many kinds of other blockchain systems have emerged, \textit{we believe these three blockchains are the most representative and popular ones, making them ideal for studying the evolution of the overall blockchain ecosystem}.

\begin{table*}[tbp]
\caption{A Comparison of Bitcoin, Ethereum and EOSIO.}
\centering
\resizebox{\textwidth}{!}{%
\begin{tabular}{lccccccc}	
\toprule
                  &\textbf{\begin{tabular}[c]{@{}c@{}}Native\\ Cryptocurrency\end{tabular}}       & \textbf{Model} & \textbf{Consensus} & \textbf{\begin{tabular}[c]{@{}c@{}}Support for\\ Smart Contract\end{tabular}} & \textbf{TPS}~\cite{garriga2020blockchain} & \textbf{Resource} &              \textbf{Role}        \\
\midrule                
\textbf{Bitcoin}   &  Bitcoin  & UTXO           & PoW                & No                                                                            & 7            & Transaction Fee   & Input / Output       \\
\textbf{Ethereum}  &  Ether    & Account/Balance        & PoW                & Yes                                                                           & 15           & Gas               & EOA / Smart Contract \\
\textbf{EOSIO}     &  EOS      & Account/Balance        & DPoS               & Yes                                                                           & 1000+         & CPU / NET         & Account             \\
\bottomrule
\end{tabular}%
}
\label{table:basic-info}
\end{table*}

\subsection{Bitcoin, Ethereum and EOSIO}
\label{sec:background:general}

We next briefly compare the working mechanisms of Bitcoin, Ethereum and EOSIO, as summarized in Table~\ref{table:basic-info}.

\subsubsection{Record-Keeping Models}
\label{sec:background:general:model}

Two types of record-keeping models are adopted in existing blockchain networks. The first method is called the UTXO (Unspent Transaction Output) Model~\cite{utxo}, and the second one is the Account/Balance Model~\cite{account-based}. Bitcoin is the only one that adopts the UTXO model, which is illustrated in Fig.~\ref{fig:utxo} in the Appendix \S\ref{sec:appendix:money-transfer}.
Specifically, each transaction is composed of a set of \textit{inputs} and \textit{outputs}, which all correspond to a certain amount of Bitcoin. The UTXO model ensures that the sum of amount represented by inputs is equal to the outputs.
Moreover, each output is locked by a \textit{scriptPubKey}. Only the one with the corresponding public key can unlock the output and spend the money by an input. Therefore, in this paper, we refer to each of the inputs and outputs as $pubkey$.

However, Bitcoin suffers from several drawbacks: 
1) the script in Bitcoin is not Turing-complete~\cite{turing}, which means it cannot perform more sophisticated functions (e.g., smart contracts); 
2) due to its consensus algorithm, the transactions per second (TPS) is low; 
and 
3) UTXOs are stateless and can be executed in parallel, which is not well suited for many applications.
Therefore, Ethereum and EOSIO adopt the Account/Balance model. It works similarly to a debit card transaction: the bank, equivalent to a consensus algorithm, tracks the balance of each debit card to make sure enough money exists before approving the transaction.

\subsubsection{Accounts and Smart Contracts}
\label{sec:background:general:account}
In Ethereum, there are two types of accounts: \textit{External Owned Accounts} (EOA) and \textit{smart contracts}.
Smart contracts contain immutable and executable bytecode, while an EOA does not and is controlled directly by a private key.
A smart contract can be created by either EOA or another smart contract. 
In contrast, the smart contract in EOSIO is \textit{updatable}. That is because an executable smart contract's bytecode in EOSIO is equivalent to an attribute of an account, like one's balance. Therefore, in EOSIO, one can easily update the smart contract in accounts he/she owns after paying the required cost. 

\subsubsection{Transactions}
\label{sec:background:general:tx}
Accounts in Ethereum and EOSIO can interact with each other by invoking \textit{transactions}, including money transfers, invoking contracts, or creating an account. 
In Ethereum, the transactions initiated by EOA and smart contracts are termed \textit{external transaction} and \textit{internal transaction}, respectively. 
In EOSIO, a transaction consists of one or more \textit{action(s)} and an action can be invoked by an account.

To better illustrate the working mechanisms, we depict the process of transferring tokens between accounts in Ethereum and EOSIO in Fig.~\ref{fig:transfer-token} (see Appendix \S\ref{sec:appendix:money-transfer}).
Specifically, when $account_a$ transfers 1 Ether to $account_b$, it initiates a transaction with an empty \textit{input} field and 1 Ether filled in the \textit{value} field. Specifically, the data in the \textit{input} field can be parsed by the recipient as a function signature with the corresponding parameters, and the \textit{value} indicates the attached Ether.
The situation in EOSIO is more complicated. 
An official smart contract, named \texttt{eosio.token}, issues EOS tokens and maintains a balance table for all holders. If $account_a$ intends to transfer EOS to $account_b$, it has to request the \texttt{eosio.token} to update the corresponding rows. After updating, the \texttt{eosio.token} would immediately \textit{notify} both of them that the transfer request is handled properly. Once $account_b$ received the notification, it would perform the following movements, e.g., initiating another action to others.
Note that, when someone intends to create an account, they need to request another official account (called \texttt{eosio}) to perform the action, instead of transferring an arbitrary amount of EOS to a non-existent address like in Ethereum.

\subsubsection{Consensus Protocol \& Resource Model}
\label{sec:background:general:consensus}
The main difference between EOSIO and the other two platforms is the consensus protocol and the resource model. PoW requires miners to calculate a string of meaningless hash values which must be preceded by a certain number of zero. Therefore, the longer the prefix of zeros, the more calculation is required, which leads to higher delays and resource-consumption.
In contrast, the DPoS is more resource–conservative and efficient. Compared to PoW, the theoretic TPS in EOSIO is hundreds-fold, reaching up to 1000+. 

Bitcoin and Ethereum adopt a similar resource model, i.e., \textit{transaction fees} and \textit{gas}. Both are used to incentivize miners to pack the transactions with the highest transaction fee or gas. 
In EOSIO, the resources are named as \textit{CPU} and \textit{NET}, which represent the allowed time to execute an action, and the space can be occupied within a single transaction. However, these two types of resource do not charge anything from initiators. They can mortgage a certain amount of EOS in exchange for \textit{CPU} and \textit{NET}, which can be reused in each transaction.
Note that, the price of CPU fluctuates according to the total value of mortgaged EOS for CPU, and NET can be freely used if some is available. 
In other words, the resource price in EOSIO is \textit{nearly free in normal conditions}.

\section{Data Collection}
\label{sec:study-design}

\subsection{Methodology}
\label{sec:study-design:data-collection}

To obtain all transactions across Bitcoin, Ethereum, and EOSIO, we deploy their official clients to synchronize data, i.e., \textit{Bitcoin Core}~\cite{bitcoin-core}, \textit{OpenEthereum}~\cite{openethereum}, and \textit{nodeos}~\cite{nodeos}. 
To conduct a fine-grained analysis, we also collect all \textit{traces}, i.e., internal transactions in Ethereum and actions in EOSIO (see \S\ref{sec:background:general:tx}). 
As the traces are too large to crawl directly from mainstream browsers (like \textit{etherscan}~\cite{etherscan}), we collect them by replaying each transactions. Specifically, \textit{trace\_transaction} in the trace module of OpenEthereum takes a transaction as an input, and replays it to collect all internal transactions that result from the given transaction. 
For obtaining actions in EOSIO, we take advantage of the approach proposed by \cite{huang2020understanding} to customize the core service daemon (nodeos) to accelerate the data collection process.

To facilitate our analysis of major events across the blockchains (see \S\ref{sec:abnormal-behaviors}), we further crawl the available information (i.e, name, category, and corresponding addresses or accounts) of all DApps from three well-known platforms: DAppTotal~\cite{dapptotal}, DappRadar~\cite{dappradar}, and DappReview~\cite{dappreview}. 

\begin{table}[t]
\caption{Dataset overview (as of March 31st, 2020).}
\centering
\resizebox{\columnwidth}{!}{%
\begin{tabular}{lccc}
\toprule
\textbf{Category} & \textbf{Bitcoin} & \textbf{Ethereum} & \textbf{EOSIO} \\
\midrule
Launch Date       & 2009.01.09     & 2015.07.30     & 2018.06.09 \\
Blocks          &  623,837     & 9,782,602 & 113,124,658 \\
\midrule
Traces            & 2,371,617,384     & 1,655,111,086     & 31,588,572,466 \\
Account           & 630,562,205      & 68,790,386        & 1,816,578      \\
Smart Contract    & -                & 24,201,516        & 6,139          \\
\midrule
DApp              & -                & 3,292             & 652            \\
DApp Addresses    & -                & 13,218            & 1,556          \\
\bottomrule
\end{tabular}
}
\label{table:dataset}
\end{table}

\subsection{Dataset Overview}
\label{sec:study-design:overview}

\subsubsection{Dataset.} In total, we have collected over \textit{35 Billion traces}. Table~\ref{table:dataset} summarizes the dataset. To the best of our knowledge, this is the \textit{largest ever} dataset studied in the research community. The \textit{trace} is the basic unit of transaction in each blockchain. For Bitcoin, a trace is an input or an output; for Ethereum, traces are composed of all the external transactions, as well as the internal ones; for EOSIO, a trace corresponds to an action.

Obviously, \textit{although EOSIO has the latest launch time, the numbers of traces and blocks are much higher than the other two blockchains.}
As Bitcoin adopted UTXO model, which does not have the concept of account, the number of Bitcoin accounts indicates the distinct \textit{pubkeys} of inputs and outputs. 
In Bitcoin, an entity may have multiple pubkeys to avoid reuse (to protect anonymity). Thus, the number of accounts in Bitcoin is tens of Ethereum's and hundreds of EOSIO's.
Moreover, we can observe a significant distinction in the number of accounts and smart contracts between Ethereum and EOSIO. As explained in \S\ref{sec:background:general:account}, this is because: 1) creating an account in EOSIO charges not only transaction fee; and 2) EOSIO smart contract is updatable. Both of them impede the explosion of the number of accounts in EOSIO.
\begin{figure}[h]
\centerline{\includegraphics[width=\columnwidth]{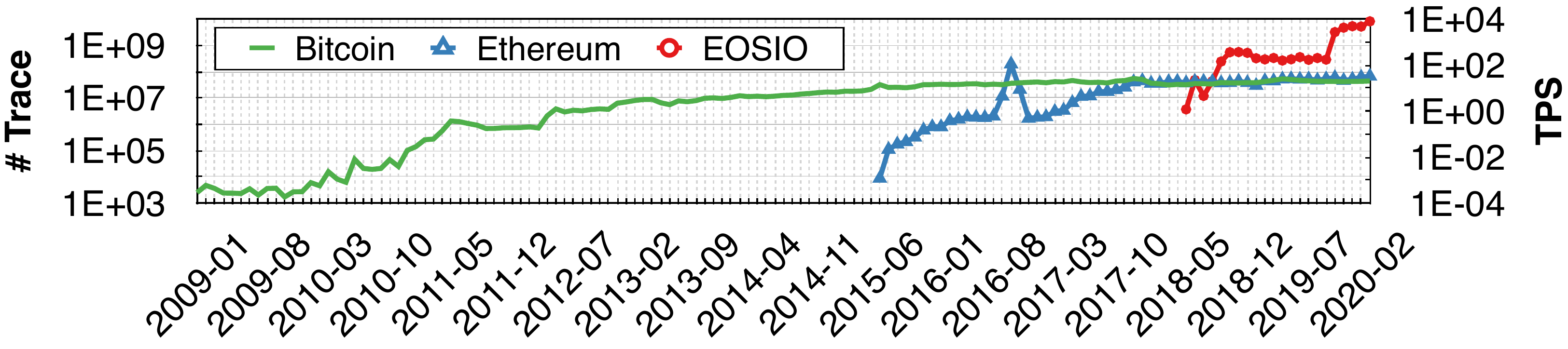}}
\caption{The evolution of traces on a monthly basis.}
\label{fig:traces}
\end{figure}

\subsubsection{General Trends.}
\label{sec:dataset:overview:general-trend}
Fig.~\ref{fig:traces} shows the number of traces collected on a monthly basis, and the corresponding Transaction per Second (TPS). 
The numbers of traces in all three platforms have an upward trend. Interestingly, it only took EOSIO three months to catch up and overtake the other two platforms. However, it took Ethereum almost three years to surpass Bitcoin.
Compared to the relatively steady growth of Bitcoin, Ethereum and EOSIO experience sudden increases. For example, in Oct. 2016, the TPS of Ethereum reached up to 73.0, around 15.9 times higher than the previous month. 
Interestingly, this explosive growth was actually due to a DoS~\cite{eth-dos} attack, which far exceeded the theoretical TPS it could handle.
As for EOSIO, the TPS reached over 1,240 in Nov. 2019, and finally reached over 3,080. This resulted from the launch of a spam DApp, which will be detailed in \S\ref{sec:abnormal-behaviors:case-study:EIDOS}.

\section{Graph-based Evolution Analysis}
\label{sec:blockchain-evolution}

\subsection{Method}
We seek to characterize the evolution of the blockchain ecosystems by investigating three representative behaviors: \textit{money transfers}, \textit{account creation}, and \textit{contract invocation}, corresponding to the types of interactions between accounts (see \S\ref{sec:background:general:tx}).
To this end, we represent these activities in three graph structures: the \textit{Money Transfer Graph (MTG)}, \textit{Account Creation Graph (ACG)}, and \textit{Contract Invocation Graph (CIG)}. 
For each blockchain platform, we will denote each graph using its first three lowercase letters in subscript, e.g., $CIG_{btc}$ for Bitcoin.
We build these three types of graphs on a monthly basis and specify the month (formatted as $yyyy-MM$) in superscript, e.g., $CIG_{btc}^{2019-10}$.
Overall, we have created 372 graphs (135 for Bitcoin, 171 for Ethereum and 66 for EOSIO).

\begin{table*}[t]
\caption{Notations with their corresponding explanations.}
\centering
\resizebox{\textwidth}{!}{%
\begin{tabular}{llll}
\toprule
\textbf{Notations} & \textbf{Explanations}            & \textbf{Notations} & \textbf{Explanations}                                                                                                                                                       \\ \midrule
$MTG$              & Money Transfer Graph             & $\alpha$           & The exponent of fitting function in degree/indegree/outdegree distribution                                                                                                  \\
$ACG$       & Account Creation Graph           & $R$                & The Pearson's Correlation Coefficient between indegree and outdegree                                                                                                        \\
$CIG$       & Contract Invocation Graph        & $WCC$, $SCC$       & Weakly and strongly connected component                                                                                                                                     \\
$T_m$              & Money transfer transactions      & $(v_i, v_j, w, t)$ & \begin{tabular}[c]{@{}l@{}}Formal definition of edges, indicating at timestamp $t$, $v_i$ initiates\\ $w$ times of corresponding type of transactions to $v_j$\end{tabular} \\
$T_a$              & Account creation transactions    & $d$                & Date, converting from timestamp $t$ and formatting as ($yyyy-MM$)                                                                                                           \\
$T_c$              & Contract invocation transactions & $V$, $E$           & The node/edge set of the graph                                          \\ \bottomrule                                                                                                   
\end{tabular}
}
\label{table:notations}
\end{table*}

\subsection{Metrics and Notations}
\label{sec:blockchain-evolution:metrics}

Following previous studies~\cite{chen2018understanding, huang2020understanding}, we measure the graphs using the following four metrics. For each graph, we later inspect each of these metrics in-turn. Table~\ref{table:notations} summarizes the key notations used in this paper.

\subsubsection{The number of traces.}
We used simplified notations $T_m$, $T_a$ and $T_c$ to refer to the \textit{money transfer traces}, \textit{account creation traces} and \textit{contract invocation traces}, respectively. \textit{This metric reflects the popularity of a certain type of activity along the timeline}. For example, a popular gambling DApp will introduce large number of $T_c$ (as the player invokes the contracts of the gambling DApp) and $T_m$ (as the player sends and receives money from the gambling DApp).
Moreover,  for each kind of trace, we tag the \textit{role of invokers}, i.e., from a smart contract or a user account. The trends can reflect whether the corresponding blockchain is mainly used as a \textit{payment network} or a \textit{decentralized system with emerging and powerful applications (i.e., smart contracts)}.

\subsubsection{Distribution of node degree.}
\label{sec:blockchain-evolution:metrics:degree}
The \textit{indegree} and \textit{outdegree} of a node represent how many edges are directed to, or pointed from that node. The \textit{degree} of a node is the sum of its indegree and outdegree. 
\textit{The node degrees distribution reflects the centralization of the blockchain in terms of certain activities.}
To measure this in our later analysis, we plot the degree/indegree/outdegree distribution for each graph, in which the y-axis is the proportion of nodes and x-axis is the size of degree/indegree/outdegree. On the basis of these plots, we take the logarithm of both axes, and perform a fit with function $y \sim x^{\alpha}$.
The \textit{the lower the $\alpha$, the shorter the tail, and the less centralized the corresponding traces}.
We illustrate a concrete example of degree distribution in \S\ref{sec:blockchain-evolution:mtg:degree}.

\subsubsection{Pearson Correlation Coefficient ($R$)}
On top of the indegree and outdegree of different nodes, we further measure the \textit{Pearson Correlation Coefficient}~\cite{pearson} (denoted as $R$ in the following) between them.
$R$ reflects whether the distribution of outdegree and indegree have a linear relationship. It is between $1$ and $-1$ to indicate strong positive or negative correlation.
For example, a popular DApp interacts frequently with users, leading to both high indegree and outdegree for all participants. To this end, the $R$ would be close to 1, indicating strong positive correlation. 
Thus, \textit{$R$ indicates the consistency of indegree and outdegree of nodes within a graph}.

\subsubsection{Weakly connected component (WCC) \& Strongly connected component (SCC)}
\label{sec:metric:wcc}
In graph theory, if there exists a \textit{path} between $v_i$ to $v_j$ in a directed graph, there exists a set of edges end-to-end whose starting and ending points are $v_i$ and $v_j$, respectively.
In WCC, there exists a path between two arbitrary nodes but without differentiating the starting point and the end point. 
In SCC, however, it requires that there exists a bi-directional path between any two nodes.
To this end, if the number of WCCs drops, it means there exists a node with extensive coverage; and if the number increases, it means such a node disappears.
Furthermore, the number of SCCs examines the possibility of two-way interaction of that node.
Consequently, \textit{the number of WCCs and SCCs reflects the connectivity between nodes in terms of one-way and two-way interaction, respectively}. For example, a spammer would initiate a large amount of $T_m$ or $T_c$ to deliver its message, but few accounts would reply. Thus, the number of WCCs and SCCs would respectively drop and rise due to many one-way edges emerging.

\section{Evolution of Money Transfers}
\label{sec:blockchain-evolution:mtg}

\subsection{Graph Construction}
\label{sec:blockchain-evolution:mtg:construction}

To build $MTG$, we first define the money transfer trace, i.e., $T_m$ in Bitcoin, Ethereum and EOSIO.
Specifically, in Ethereum, a transaction with a certain amount of Ether and a blank input field is regarded as a $T_m$. The payer and payee can be identified easily by their addresses.
In EOSIO, we further parse the following notifications initiated from \texttt{eosio.token} after transfer requests. Hence, we can extract the real participants and construct a $T_m$ directly between them.
For Bitcoin, a transaction may have multiple inputs and outputs, which are regarded as two sets: payers and payees. Therefore, 
we follow previous work~\cite{akcora2018forecasting} and insert an additional node and label it by the corresponding transaction id, $txid$.

As defined in Table~\ref{table:notations}, each $T_m$ can be formally represented as $(v_i, v_j, w, t)$, indicating at time $t$, $v_i$ transferred $w$ tokens to $v_j$.
Then, by converting the timestamp $t$ to date $d$ (formatted as $yyyy-MM$) and adding $w$ up as $w^*$, we can group $T_m$s with the same payer and payee on a monthly basis for each platform. This can be formalized as $\{(v_i, v_j, w^*, d) | v_i, v_j \in V, w^* \in \mathbb{R}^+\}$.
Consequently, $MTG$s can be defined as a \textit{weighted directed graph}: $MTG = (V, E)$, where $E$ is the above mentioned set and $V$ is a set of nodes represented as: $(id, label)$. Specifically, $id$ is the identifier to distinguish nodes within a platform; and the $label$ is the category of DApps to which the $id$ belongs, according to the off-chain information we collected.
Consequently, we have created 214 MFGs, 135 for Bitcoin, 57 for Ethereum and 22 for EOSIO.
We spend the rest of this section inspecting $MTG$ using the four metrics defined in \S\ref{sec:blockchain-evolution}.

\subsection{Number of Traces}
\label{sec:blockchain-evolution:mtg:traces}

\subsubsection{Overall evolution.}
\begin{figure}[t]
\centerline{\includegraphics[width=1\columnwidth]{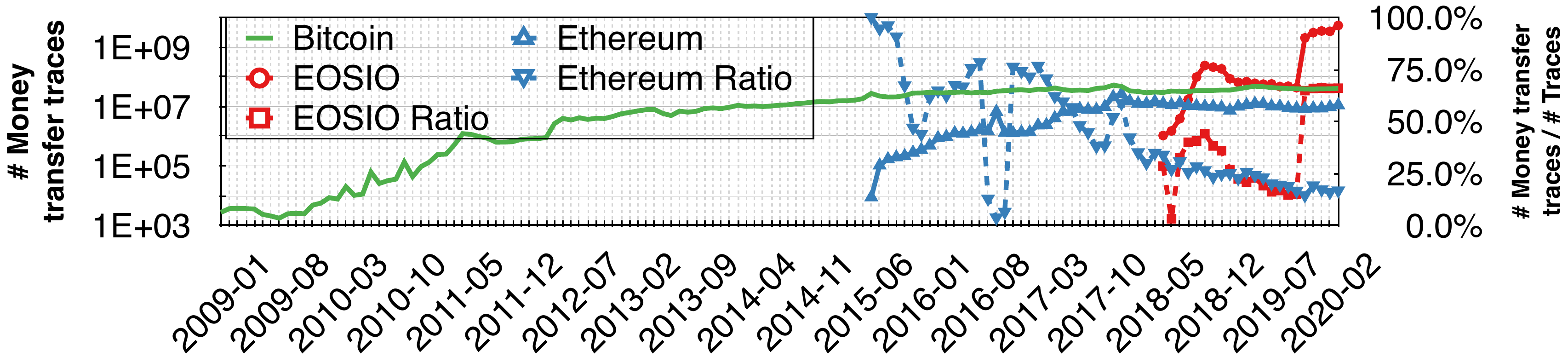}}
\caption{The evolution of monthly money transfer traces across three blockchains.}
\label{fig:traces-mtg}
\end{figure}

The number of $T_m$ over time across three platforms is shown in Fig.~\ref{fig:traces-mtg}.
The number of $T_m$ in Bitcoin has always been greater than Ethereum's.
Since Oct. 2018, the number of $T_m$ in EOSIO has surpassed the other two platforms. Though there was a decline from late 2018, its number has reached 2.1 billion in $MTG_{eos}^{2019-11}$, which is more than an order of magnitude higher than Bitcoin and Ethereum at the same time.

We further show the proportion of $T_m$ out of all traces by a dashed line in Fig.~\ref{fig:traces-mtg}. Note that we omit the percentage of Bitcoin, as it is always 100\% (the Bitcoin network is solely composed of $T_m$).
For Ethereum, we observe three troughs in late-2015, Oct. 2016 and Nov. 2017. These were caused by Dwarfpool~\cite{dwarf}, a DoS attack~\cite{eth-dos}, and a DApp called CryptoKitties~\cite{cryptokitties}, respectively. 
To be specific, the mining pool and the DoS attack both invoked a huge number of $T_c$ (see \S\ref{sec:blockchain-evolution:cig}). The DoS attack took advantage of the flawed design of the price of gas (see \S\ref{sec:background:general:consensus}).
CryptoKitties is a collectible game where players can trade and exchange \textit{kitties}. However, \textit{kitties} are represented by alternative tokens, whose transfer is regarded as $T_c$ in our paper.
Except for these three troughs, the overall share of $T_m$ in Ethereum has been gradually decreasing, which we believe is due to the emergence of DeFi since May 2017. According to our data, in just six months, the number of DeFi related $T_c$ had risen by a remarkable 486.3\% and another 7,420.4\% for the next half year.

In contrast, the amount of $T_m$ in EOSIO surged in late-2018, which was due to the popularity of gambling DApps. This type of DApps requires frequent and abundant money transferred between players and contracts. Our collected data indicates there were 237 million $T_m$ related to gambling DApps in EOSIO in Nov. 2018 (equivalent to 23.4 times of all $T_m$ in Ethereum in the same month).
After this, the share of $T_m$ also started to go down until Nov. 2019. The emergence of EIDOS\footnote{This accepts a user's transfer of EOS and returns it back immediately. In the meanwhile, it initiates another transaction with a certain amount of EIDOS tokens, issued by themselves but can be traded in exchanges. Therefore, the ratio of $T_m$ to $T_c$ related to EIDOS is around 2 to 1.} \cite{eidos} forced the percentage of $T_m$ to around 66\% (i.e. two-thirds) of all traces in each month (see \S\ref{sec:abnormal-behaviors:case-study:EIDOS}).

\begin{figure}[h]
     \centering
     \begin{subfigure}[t]{0.8\columnwidth}
         \centering
         \includegraphics[width=\textwidth]{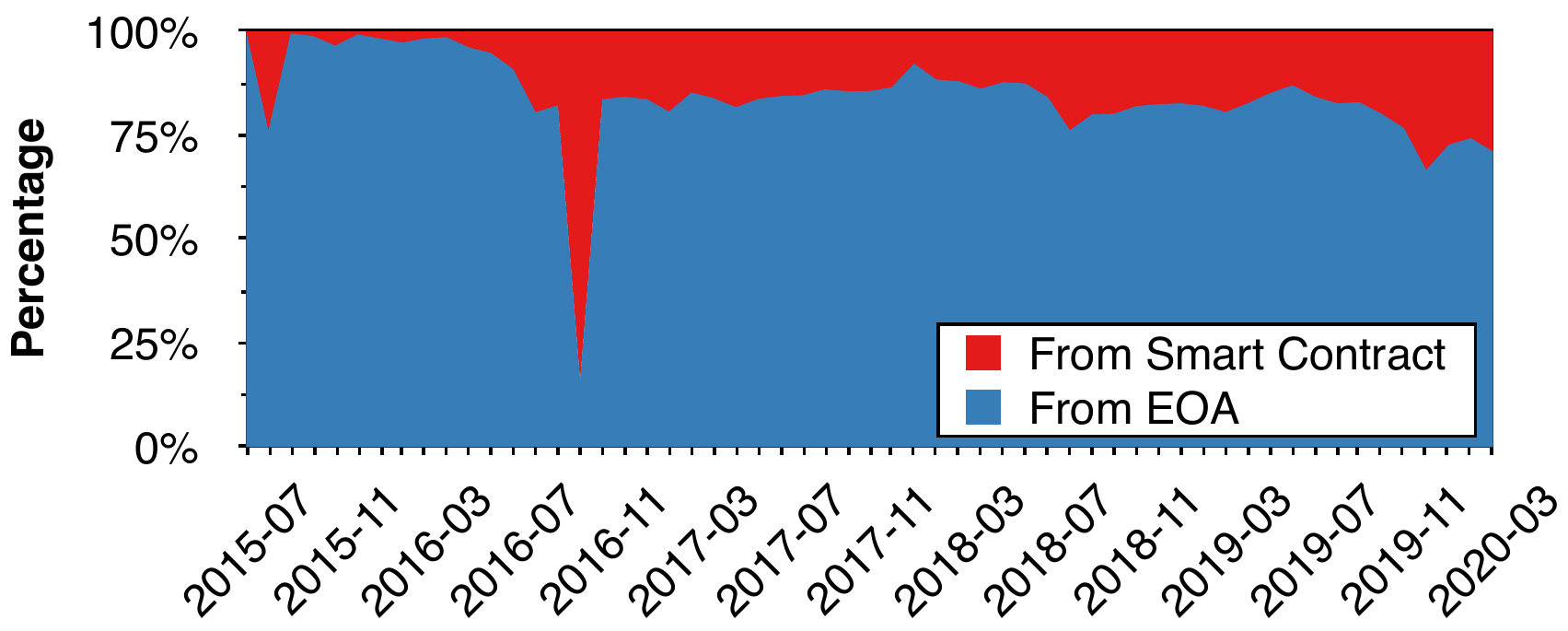}
         \caption{Ethereum}
         \label{fig:percentage-mtg-eth}
     \end{subfigure}
     \begin{subfigure}[t]{0.8\columnwidth}
         \centering
         \includegraphics[width=\textwidth]{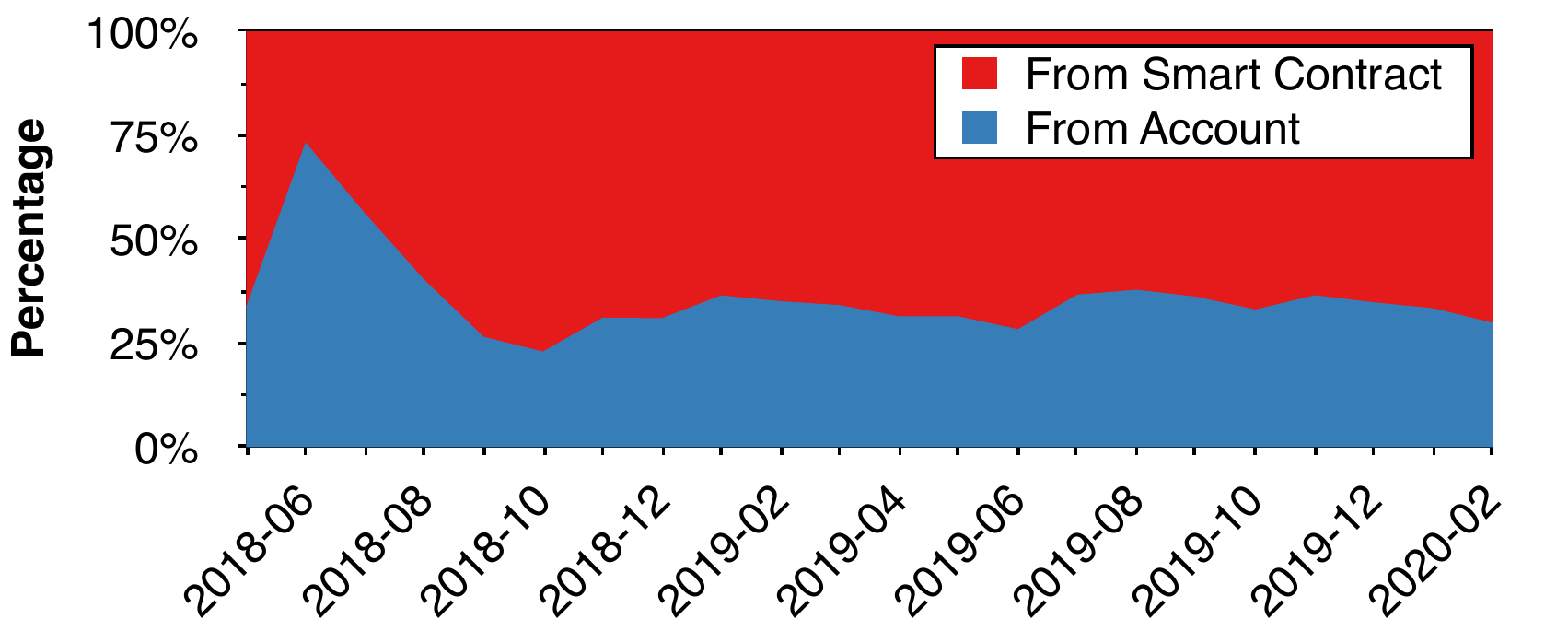}
         \caption{EOSIO}
         \label{fig:percentage-mtg-eos}
     \end{subfigure}
        \caption{Percentage of initiators' role who invokes money transfer traces.}
        \label{fig:percentage-mtg}
\end{figure}

\subsubsection{The role of initiators.}
Fig.~\ref{fig:percentage-mtg} shows the percentage of the role of initiators, from whom the $T_m$ is initiated, in Ethereum and EOSIO.
An obvious trough existed in Oct. 2016 from Fig.~\ref{fig:percentage-mtg-eth}. 
This is because the DoS attack requires frequent invocations of $T_m$, which can be easily achieved by smart contracts instead of regular accounts (i.e., human).
Except for that, we observe that users in Ethereum prefer transferring money on their own, though the ratio was still gradually going down.
According to our data, along with the increase of the absolute number of $T_m$ (see Fig.~\ref{fig:traces-mtg}), since 2018, the number of $T_m$ initiated by EOA has dropped by 67.6\%.
Moreover, in EOSIO, we see a surge of $T_m$ initiated from smart contracts in late-2018, as the contracts of gambling DApps needed to invoke $T_m$ to return money to their winners. Our data illustrates that among these 237M pieces of gambling related $T_m$, 78.38\% of them were initiated by gambling DApps.

\textbf{\textit{Insight: }}
\textit{
The activity of the blockchain ecosystems has shown continuous growth in terms of transferring native tokens, in-part incentivized by killer DApps.
The proportion of transfer transactions reveals that both Ethereum and EOSIO has been experiencing a decline, due to them departing from pure value-transfer networks.
}

\subsection{Degree}
\label{sec:blockchain-evolution:mtg:degree}
\begin{figure}[h]
     \centering
     \begin{subfigure}[t]{0.4\columnwidth}
         \centering
         \includegraphics[width=\textwidth]{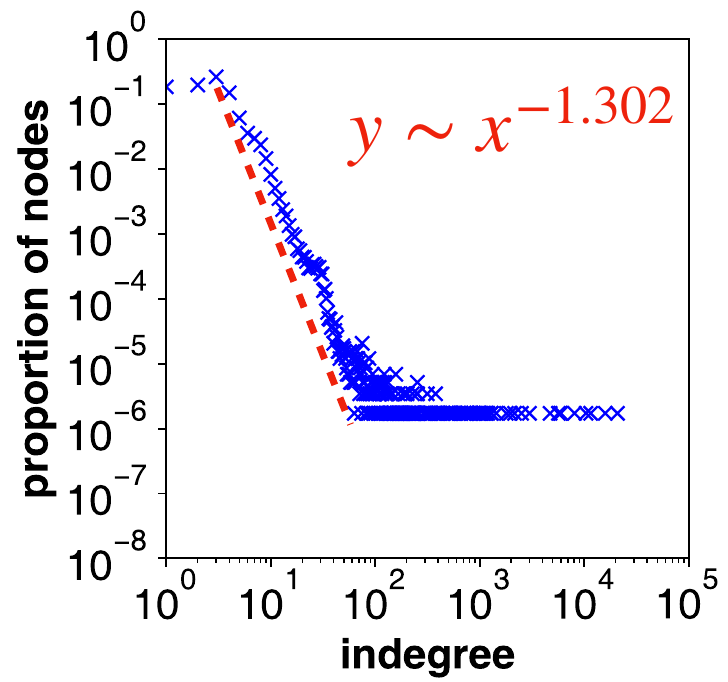}
         \caption{EOSIO indegree}
         \label{fig:alpha-eg-eos-in}
     \end{subfigure}
     \begin{subfigure}[t]{0.4\columnwidth}
         \centering
         \includegraphics[width=\textwidth]{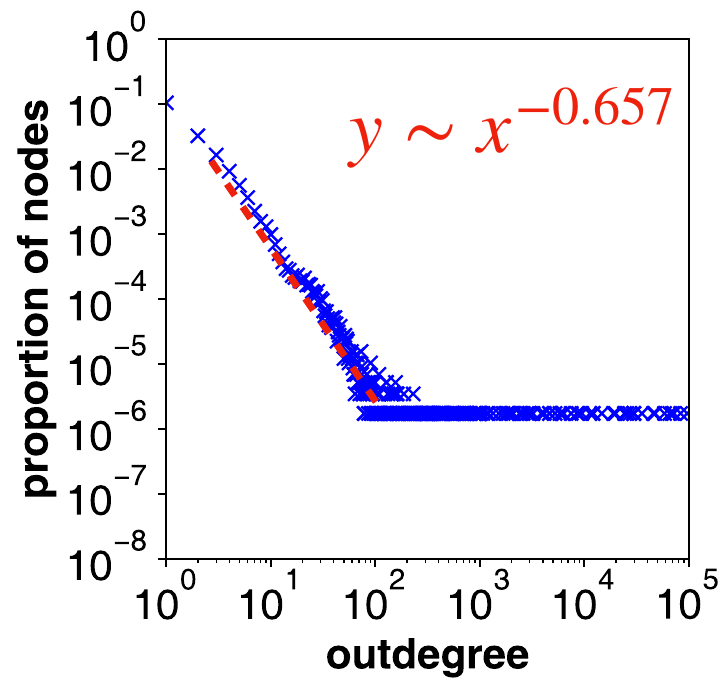}
         \caption{EOSIO outdegree}
         \label{fig:alpha-eg-eos-out}
     \end{subfigure}
     \begin{subfigure}[t]{0.4\columnwidth}
         \centering
         \includegraphics[width=\textwidth]{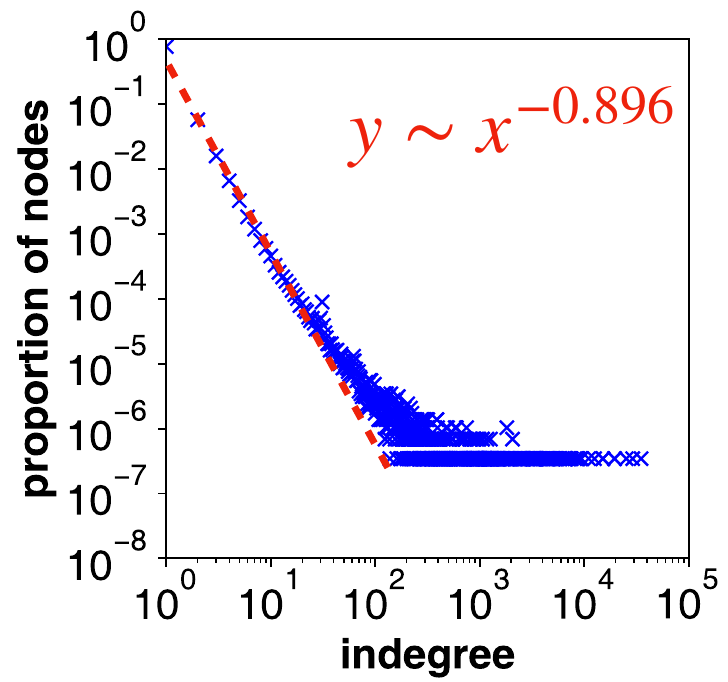}
         \caption{Ethereum indegree}
         \label{fig:alpha-eg-eth-in}
     \end{subfigure}
     \begin{subfigure}[t]{0.4\columnwidth}
         \centering
         \includegraphics[width=\textwidth]{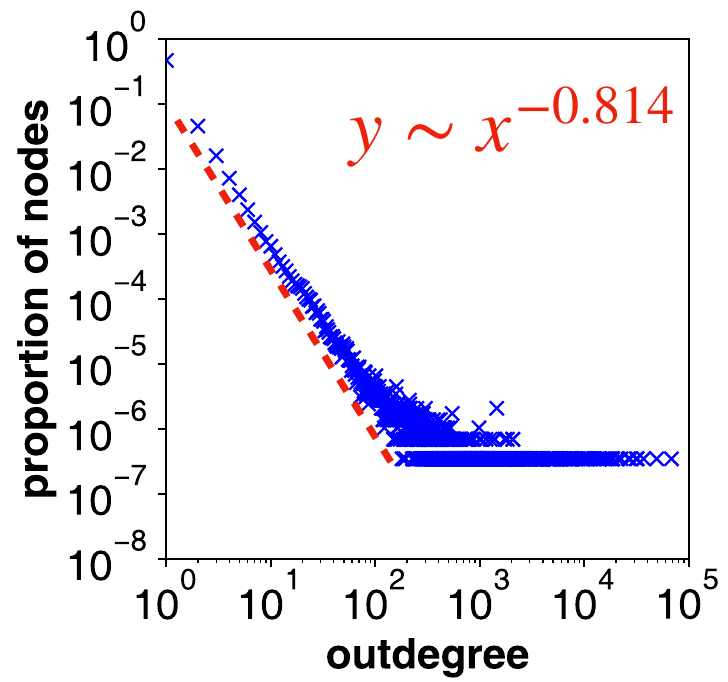}
         \caption{Ethereum outdegree}
         \label{fig:alpha-eg-eth-out}
     \end{subfigure}
        \caption{Indegree and outdegree distribution for $MTG_{eos}^{2019-10}$ and $MTG_{eth}^{2019-10}$.}
        \label{fig:alpha-eg}
\end{figure}

As explained in \S\ref{sec:blockchain-evolution:metrics}, we use $\alpha$ to measure the degree distribution. 
We first illustrate its meaning using a concrete example. 
Fig.~\ref{fig:alpha-eg} illustrates the $\alpha$ of indegree and outdegree distribution in $MTG_{eos}^{2019-10}$ and $MTG_{eth}^{2019-10}$.
Each dot in Fig.~\ref{fig:alpha-eg} shows that there is $y$ percent of nodes with the degree/indegree/outdegree valued as $x$ (in the corresponding month and platform). 
The distribution of node degree is quite different across the blockchains.
Note, the larger the $\alpha$, the more obvious the long-tail distribution, indicating less variabilty in the nodes' degree. 
More importantly, a high $\alpha$ means greater centralization.
For example, we identify a family of accounts in EOSIO that invoke spam advertisements in Oct. 2019 (more details in \S\ref{sec:abnormal-behaviors:case-study:spam}).
Due to this, Fig.~\ref{fig:alpha-eg-eos-in} has points in the top left corner. These are nodes with a small indegree, representing victims of this spam attack. In contrast, we see a set of points located in the bottom right corner of Fig.~\ref{fig:alpha-eg-eos-out}, i.e., the spammers.

\subsubsection{Overall evolution.}
\begin{figure}[h]
     \centering
     \begin{subfigure}[t]{\columnwidth}
         \centering
         \includegraphics[width=\textwidth]{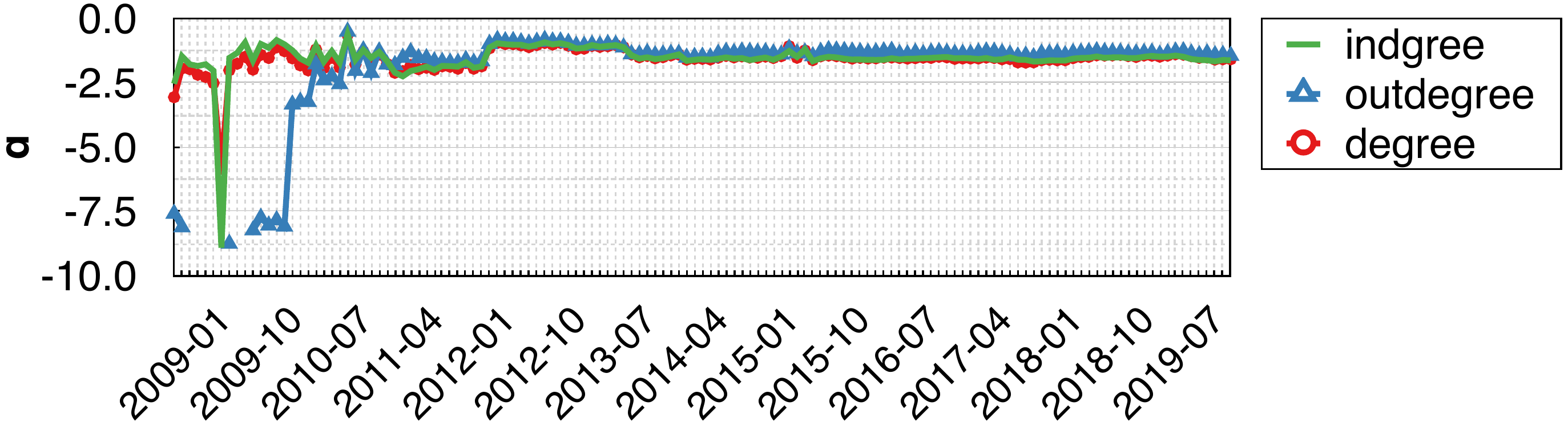}
         \caption{Bitcoin}
         \label{fig:alpha-mtg-btc}
     \end{subfigure}
     \begin{subfigure}[t]{0.63\columnwidth}
         \centering
         \includegraphics[width=\textwidth]{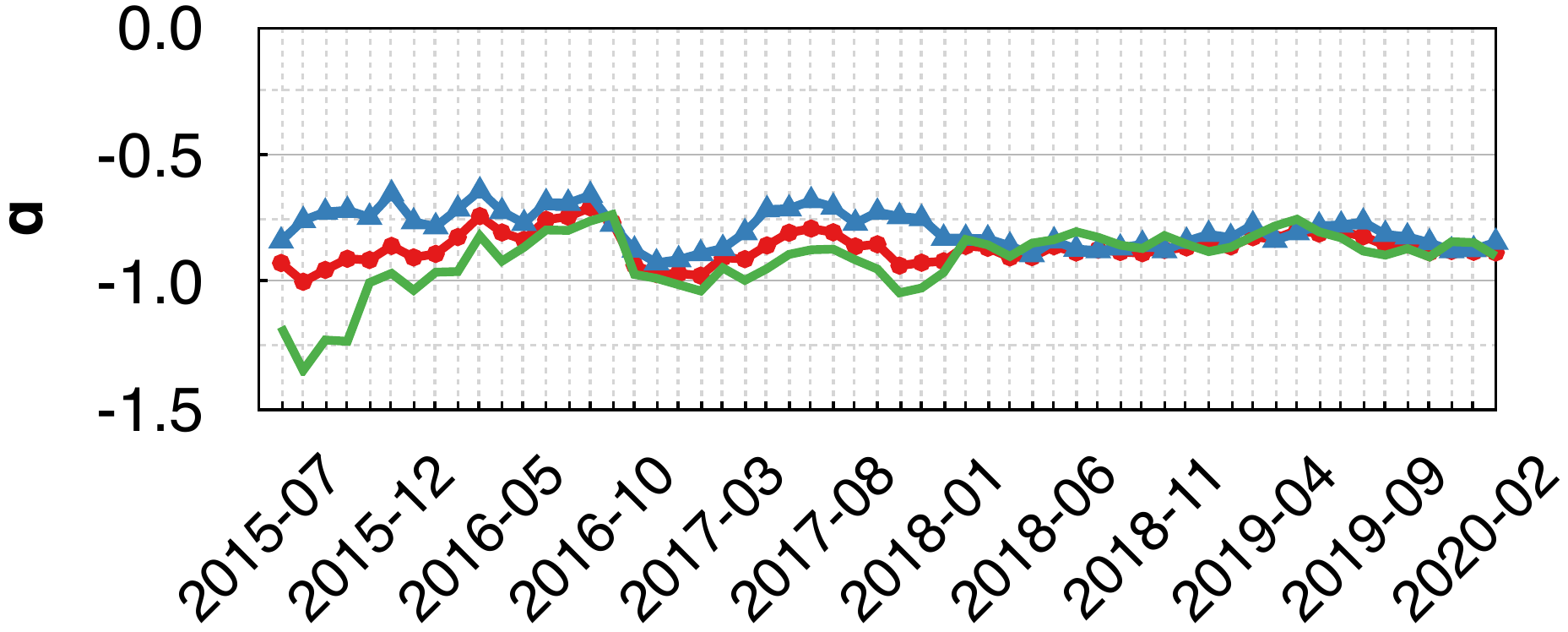}
         \caption{Ethereum}
         \label{fig:alpha-mtg-eth}
     \end{subfigure}
     \begin{subfigure}[t]{0.36\columnwidth}
         \centering
         \includegraphics[width=\textwidth]{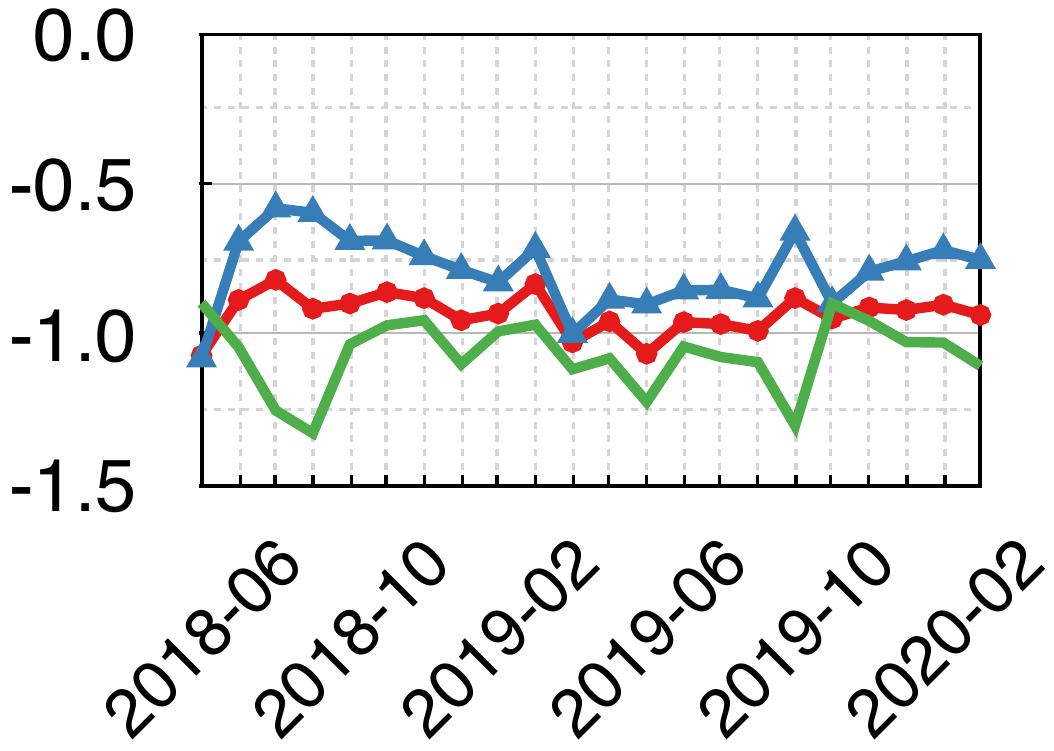}
         \caption{EOSIO}
         \label{fig:alpha-mtg-eos}
     \end{subfigure}
        \caption{$\alpha$ of degree/indegree/outdegree distribution of blockchain platforms over time in terms of $MTG$.}
        \label{fig:alpha-mtg}
\end{figure}

To study this centralization, we further calculate the $\alpha$ of the degree/indegree/outdegree distribution for each platform on a monthly basis. The trend of the $\alpha$s are shown in Fig.~\ref{fig:alpha-mtg}.
The general trend shows that, in Bitcoin, except for the infant stage (prior to 2012), the $\alpha$ of indegree and outdegree distributions were quite stable and changed almost simultaneously.
Moreover, the $\alpha$ of Ethereum's indegree distribution shows an overall gradual upward trend, while that of outdegree was gradually stabilizing. This signals growing centralization of indegree distribution.
In EOSIO, the $\alpha$ metrics fluctuate significantly though. Compared to Ethereum, the indegree was less centralized but the outdegree was exactly in the opposite position.

\subsubsection{Per-chain analysis.}
When we look in detail on the per-chain basis, subtle differences arise.
As we see from Fig.~\ref{fig:alpha-mtg-btc}, prior to 2012, 
Bitcoin was still in its infancy and the number of $T_m$ per month was so small that a few transactions could cause a large shift in the $\alpha$.
From Fig.~\ref{fig:alpha-mtg-eth}, we see that prior to mid-2018, the $\alpha$ of outdegree distribution went higher than that of the indegree distribution. Combining with the historical prices of Ether~\cite{eth-cap}, we see that the price was going up and reached a local maximum from mid-2017. This would motivate frequent buy and sell transactions through exchanges, who typically have multiple entries used to collect deposit requests from users. Hence, this has resulted in more centralized indegree distribution than outdegree. According to our data, the exchange related $T_m$ during late 2017 accounted for 15.7\% --- 48.18\% of the total DApp-related $T_m$.
For EOSIO,  except for Oct. 2019, we see the trend of $\alpha$ was similar with the Ethereum's (see Fig.~\ref{fig:alpha-mtg-eos},). Also, if we combine the historical price of EOS~\cite{eos-cap}, we see that there were some price peaks, especially before Dec. 2018 and Jun. 2019.
However, the existence of exchanges in EOSIO cannot drive the $\alpha$ so dramatically. 
Other driving forces include the growth of gambling DApps in Sep. 2018,
i.e., around 6.9 million transactions were initiated from gambling DApps. The top 1\% accounted for 98.1\% of them, which explains the  greater centralization of outdegree.

\textbf{\textit{Insight: }}
\textit{
Compared to the relative stable degree distribution in Bitcoin, the degree distribution in Ethereum and EOSIO has been significantly affected by the price of tokens. This has resulted in greater centralization (with some dominant accounts) that are reflected in the fluctuation of $\alpha$.
}

\subsection{Pearson Correlation Coefficient ($R$)}
\label{sec:blockchain-evolution:mtg:pearson}

\begin{figure}[h]
\centerline{\includegraphics[width=\columnwidth]{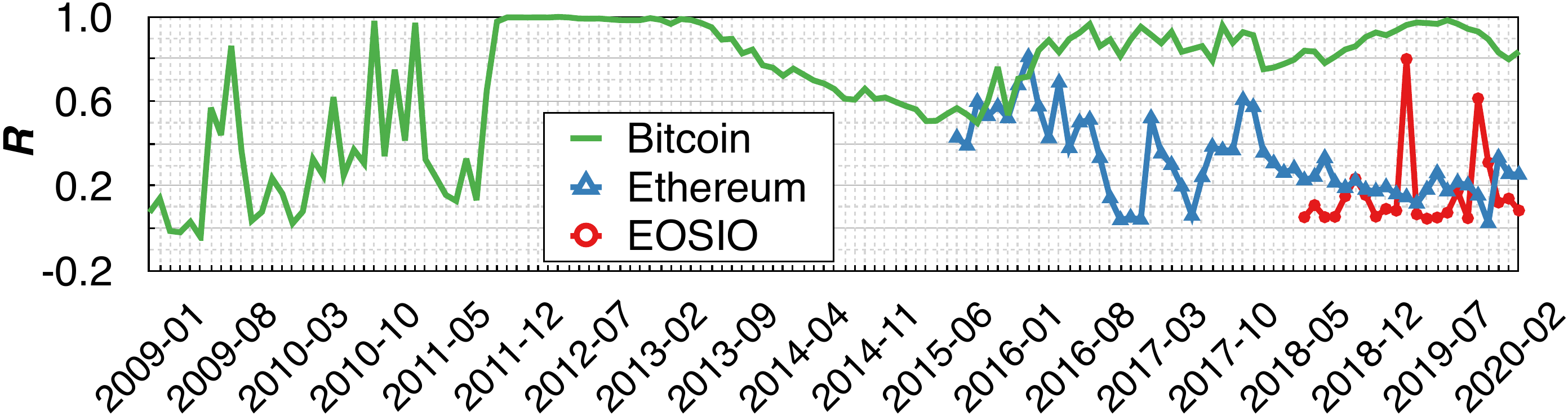}}
\caption{Pearson's correlation between indegree and outdegree of blockchain platforms over time.}
\label{fig:pearson-mtg}
\end{figure}

The correlation ($R$) between indegree and outdegree of nodes for these platforms are depicted in Fig.~\ref{fig:pearson-mtg}.
After the infant stage of Bitcoin,
its $R$ was relatively high, reflecting a strong positive correlation of indegree and outdegree. However, this may be due to the topology of $MTG_{btc}$, i.e., the node
must have one indegree (the referring output is directed from its $txid$) and zero/one outdegree (depending on if it is (un)spent).
In Ethereum, after the DoS attack, there were two peaks that indicate a strong correlation: Mar. 2017 (0.521) and Dec. 2017 (0.607). These are caused by ShapeShift~\cite{shapeshift} and Cryptokitties, respectively. The former is an exchange which allows users to exchange cryptocurrencies. Though Cryptokitties led to frequent token exchanges, it still required a huge amount of Ether transfer to perform game logic, like trading and breeding \textit{kitties}. In terms of $T_m$, the core contract of Cryptokitties accounted for 469K and 458K outdegree and indegree, respectively.
However, this has been gradually decreasing since 2018, due to the emergence of DeFi, which is more likely to initiate $T_c$ instead of $T_m$ (see \S\ref{sec:blockchain-evolution:cig}).

As for EOSIO, the overall tendency of $R$ shows there was nearly no correlation between indegree and outdegree, except for two months: Apr. 2019 and Nov. 2019. One previous work~\cite{huang2020understanding} analyzed the EOS Global event that happened in Apr. 2019. This was a bot-like account that sent and received a huge number of EOS to fake an exaggerated trading volume. This resulted in huge outdegree and indegree.
The peak in Nov. 2019 resulted from EIDOS, a spam airdrop event, whose contract received a lot of EOS from investors and returned them back immediately (detailed in \S\ref{sec:abnormal-behaviors:case-study:EIDOS}).

\textbf{\textit{Insight: }}
\textit{
Bitcoin has a relatively high $R$, but Ethereum's $R$ has been gradually decreasing. This may be related to the emergence of popoular DeFi DApps.
In contrast, EOSIO has maintained a relatively low level of $R$ in transferring money, except for the peaks imported by exchanges and spams.
}

\subsection{Connected Components}
\label{sec:blockchain-evolution:mtg:cc}
\begin{figure}[t]
\centerline{\includegraphics[width=\columnwidth]{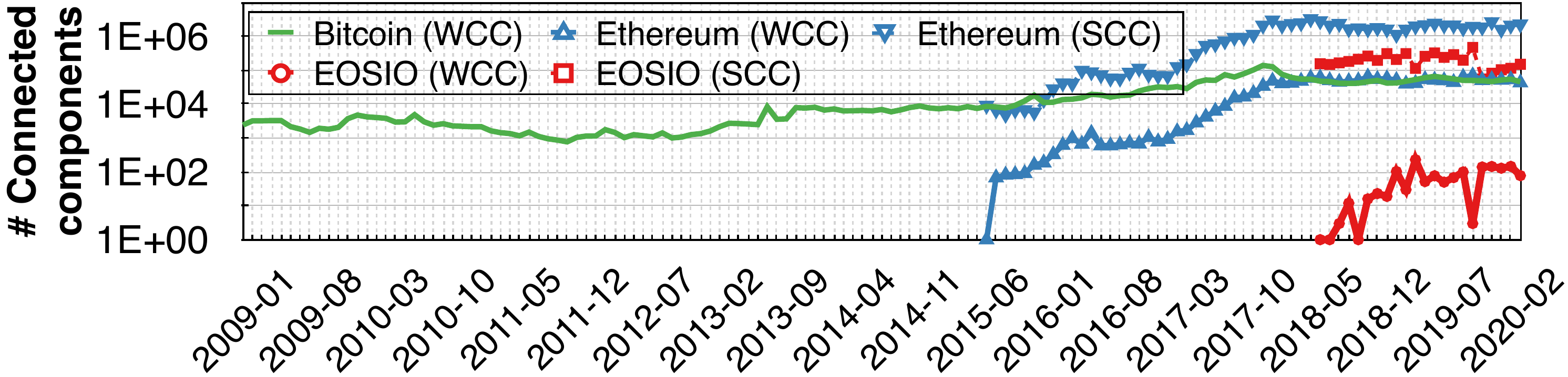}}
\caption{The number of WCC and SCC in different blockchain platforms over time.}
\label{fig:cc-mtg}
\end{figure}

Fig.~\ref{fig:cc-mtg} shows the evolution of the numbers of WCC and SCC.
Generally, these numbers reflect the (dis)apperance of nodes who could import huge amounts of (bi-)directional edges (see \S\ref{sec:metric:wcc}).
Note that, due to the impossibility of constructing bi-directional edges in $MTG_{btc}$ for the Bitcoin network, we only measure its number of WCC.
Until Apr. 2018, the number of WCCs for Ethereum and Bitcoin have been almost equal (48,529 for Ethereum and 55,387 for Bitcoin). In contrast, the number of WCC for EOSIO was only 230 at most (in Apr. 2019). There are two possible reasons: 
1) a relatively lower number of accounts and smart contracts in EOSIO compared to Ethereum and Bitcoin (see Table~\ref{table:dataset}); 
and 2) $T_m$ is more likely to be invoked in EOSIO (see Fig.~\ref{fig:traces-mtg}).
In Oct. 2019, the number of WCC in EOSIO dropped to 3. This was caused by an enormous amount of spam advertisements (detailed in \S\ref{sec:abnormal-behaviors:case-study:spam}) that carry tiny EOS but cover many accounts. Astonishingly, this pushes the size of WCC as 579,964, covering more than 99.9\% of nodes in $MTG_{eos}^{2019-10}$.

As to SCC, the difference between EOSIO and Ethereum has reduced, which suggests that one-way directed edges in $T_m$ still dominate. In EOSIO, we also see a sudden drop in Nov. 2019 (the number of SCC is 55,452), as the EIDOS enforces the bi-directional $T_m$.

\textbf{\textit{Insight: }}
\textit{The number of WCC was relatively low in EOSIO compared to other two platforms, suggesting that accounts in EOSIO have stronger connections than those in Bitcoin and Ethereum. That said, violent fluctuations were introduced by certain misbehaviors.
Moreover, the numbers of WCCs and SCCs have an upward trend, meaning that the platforms are constantly flooded with new users. However, our results show that the existing contracts do not perform many money transfers with them. 
}

\section{Evolution of Account Creation}
\label{sec:blockchain-evolution:acg}

\subsection{Graph Construction}
Though we uniformly defined accounts in all three platforms (see \S\ref{sec:study-design:overview}), the concept of an `accounts' in Bitcoin is still vague. Therefore, we only discuss $T_a$, \textit{account creation transactions}, in Ethereum and EOSIO.
In Ethereum, we focus on the smart contract, which can be created by an internal transaction whose type is \textit{create} (see \S\ref{sec:background:general:tx}).
As for EOSIO, we parse the notifications followed by function \texttt{newaccount} in theofficial account \texttt{eosio} to parse the actual creator and the created account (explained in \S\ref{sec:background:general:tx}).

Similar to the definition of $T_m$ (see \S\ref{sec:blockchain-evolution:mtg:construction} and Table~\ref{table:notations}), $T_a$ is also defined as $(v_i, v_j, w, t)$. However, as each $v_j$ can only be created once, the $w$ is always $1$ here.
Therefore, we can group $T_a$s by only converting $t$ as $d$: $\{(v_i, v_j, d) | v_i, v_j \in V\}$.
Eventually, we obtain an \textit{unweighted directed graph} $ACG$ for each month: $ACG = (V, E)$, where $E$ is the above set and $V$ is the set composed of nodes as introduced in \S\ref{sec:blockchain-evolution:mtg:construction}.
In total, we have created 79 ACGs: 57 for Ethereum and 22 for EOSIO.
We spend the rest of this section inspecting $ACG$ using three of the metrics defined in \S\ref{sec:blockchain-evolution}.

\begin{figure}[t]
\centerline{\includegraphics[width=\columnwidth]{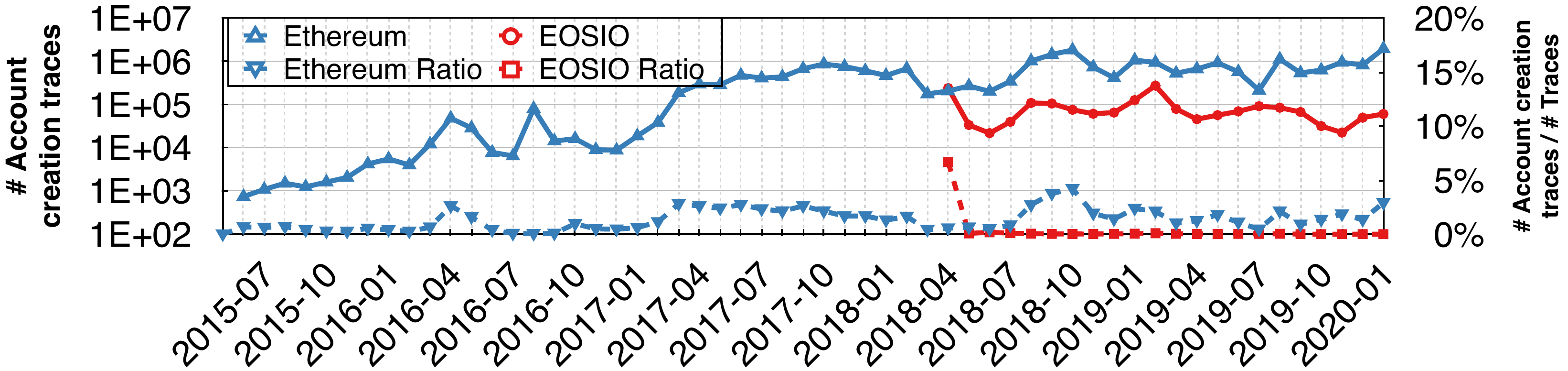}}
\caption{The evolution of account creation traces.}
\label{fig:traces-acg}
\end{figure}

\begin{figure}[h]
     \centering
     \begin{subfigure}[t]{0.8\columnwidth}
         \centering
         \includegraphics[width=\textwidth]{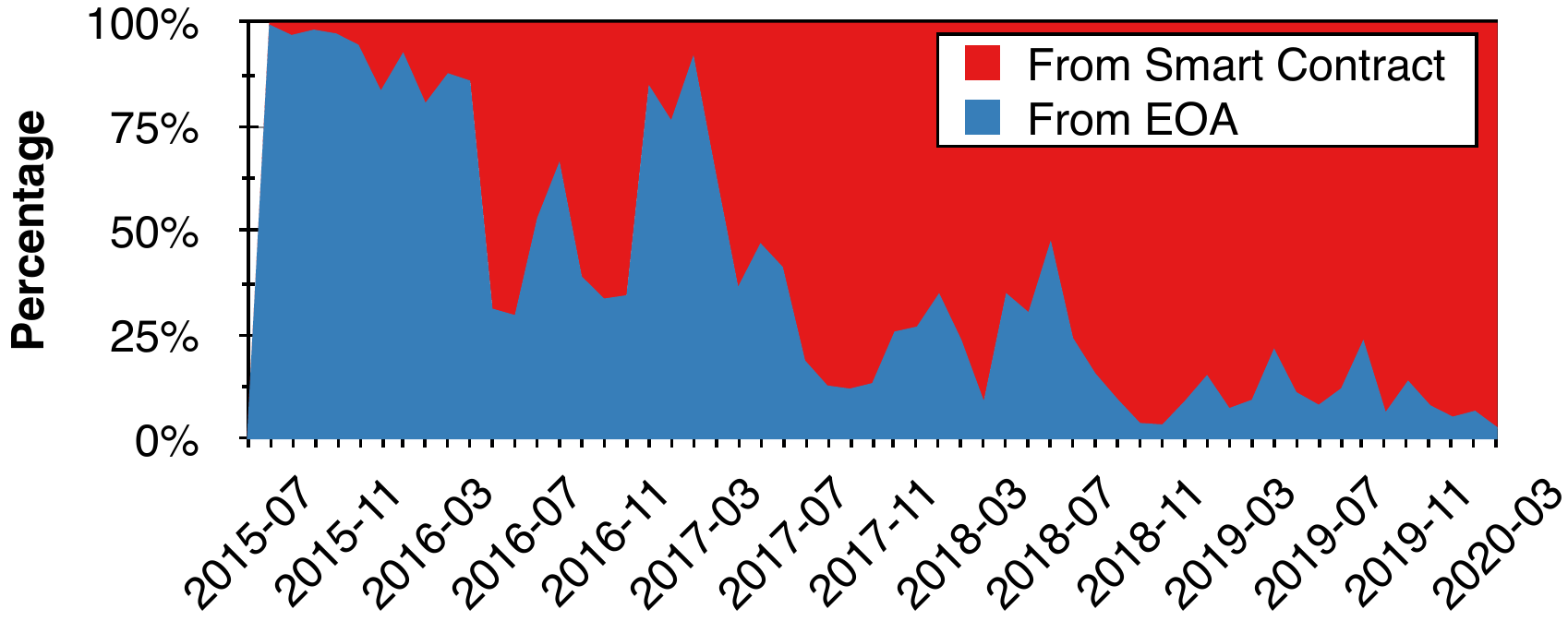}
         \caption{Ethereum}
         \label{fig:percentage-acg-eth}
     \end{subfigure}
     \begin{subfigure}[t]{0.8\columnwidth}
         \centering
         \includegraphics[width=\textwidth]{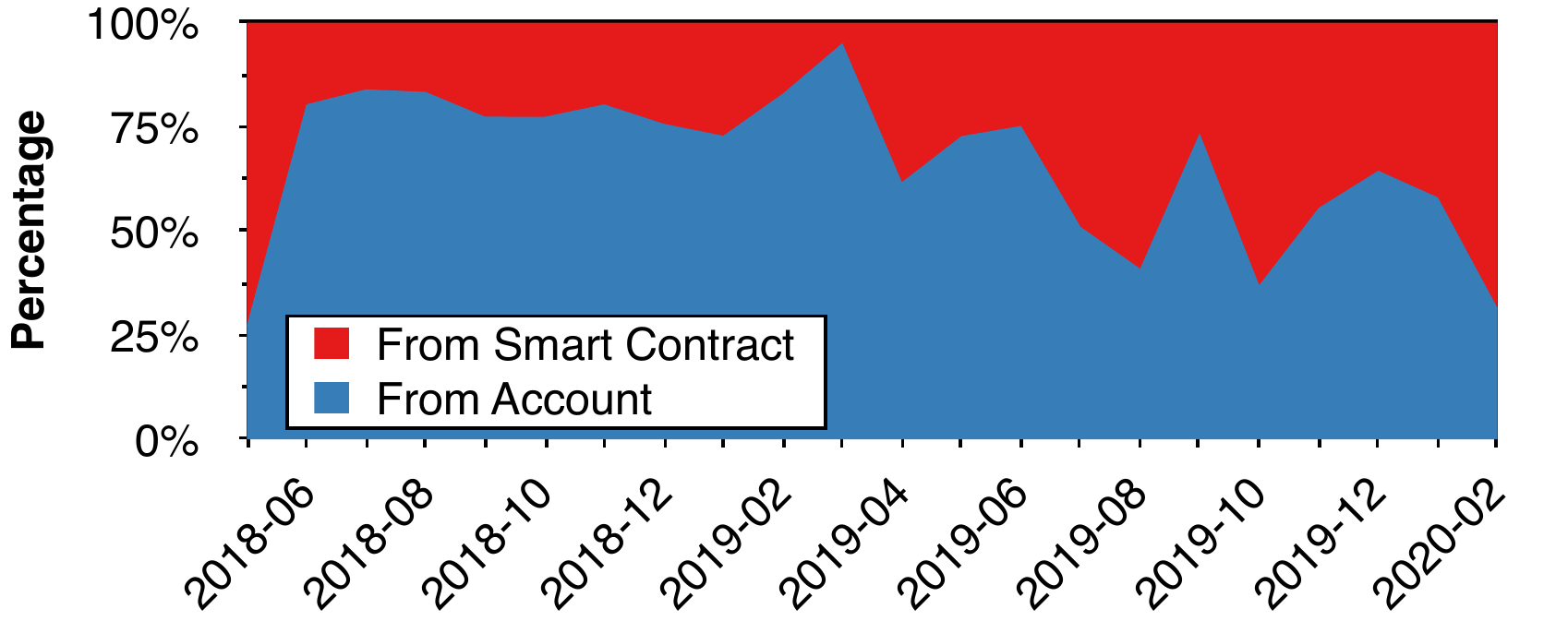}
         \caption{EOSIO}
         \label{fig:percentage-acg-eos}
     \end{subfigure}
        \caption{The distribution of initiators of $T_a$.}
        \label{fig:percentage-acg}
\end{figure}

\subsection{Number of Traces}
\label{sec:blockchain-evolution:acg:traces}
The number of $T_a$ and its corresponding proportion to the number of all traces in the corresponding month are shown in Fig.~\ref{fig:traces-acg}.
Taking the number of $T_a$ in Ethereum as a whole, we can observe a conspicuous upwards trend. Contrary to Ethereum, the number of newly created accounts in EOSIO has remained at a relatively low but stable level. The exception is during Apr. 2019, where the EOS Global contract created more than 241K bot-accounts.
Moreover, in EOSIO, except for the first month, the ratio of $T_a$ was almost negligible (0.17\% in Aug. 2018 was the maximum ratio).
Contrary, there was always a certain percentage of $T_a$ in Ethereum for each month (around 1\% -- 2.5\%), though the percentage was much lower than $T_m$'s (see Fig.~\ref{fig:traces-mtg}).
The main reason is that creating an account in Ethereum is almost free compared to EOSIO\footnote{
Creating an EOSIO account will cost 5.39 USD~\cite{eos-create-account}, while it takes less than 0.01 USD in Ethereum on that day.
}, which may make users more reluctant to reuse existing accounts for different purposes.

Fig.~\ref{fig:percentage-acg} further shows the role of the initiator who invokes $T_a$s.
Especially in Mar. 2020, the ratio of Ethereum accounts created by smart contracts is 33.7 times higher than that of EOAs. Such an astonishing ratio shows a dramatic contrast to the ratio in $MTG$ (see Fig.~\ref{fig:percentage-mtg-eth}).
After manual investigation, we find that this is due to a contract named GasToken~\cite{gastoken}, which helps users to spend less gas to complete the transaction --- this created a huge amount of accounts since 2019. 
The $T_a$ it initiated accounted for 62\% of all $T_a$ in Mar. 2020.
Moreover, we find that the ratio of $T_a$ initiated from contracts has little correlation with the prosperity of DApps.
Overall, the DApp related account creation transactions has never accounted for higher than 15\% in each month.

\textbf{\textit{Insight: }}
\textit{In Ethereum and EOSIO, the absolute number and ratio of account creation transactions are relative low compared to transfer transactions. Creating accounts is less necessary in Ethereum, and EOSIO charges expensive fees.
Moreover, accounts are more likely to be created by contracts, indicating the tendency for more relationships between smart contract.
}

\subsection{Degree}
\label{sec:blockchain-evolution:acg:degree}
\begin{figure}[t]
     \centering
     \begin{subfigure}[t]{0.63\columnwidth}
         \centering
         \includegraphics[width=\textwidth]{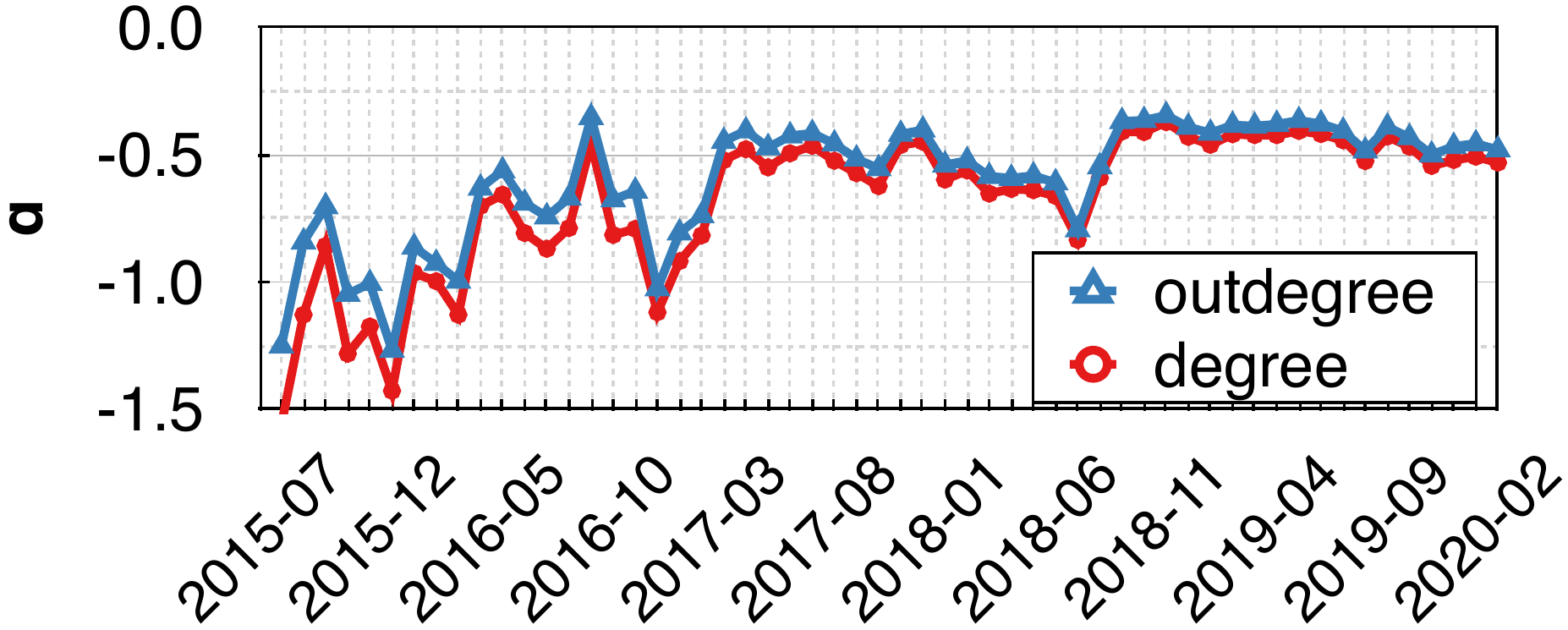}
         \caption{Ethereum}
         \label{fig:alpha-acg-eth}
     \end{subfigure}
     \begin{subfigure}[t]{0.36\columnwidth}
         \centering
         \includegraphics[width=\textwidth]{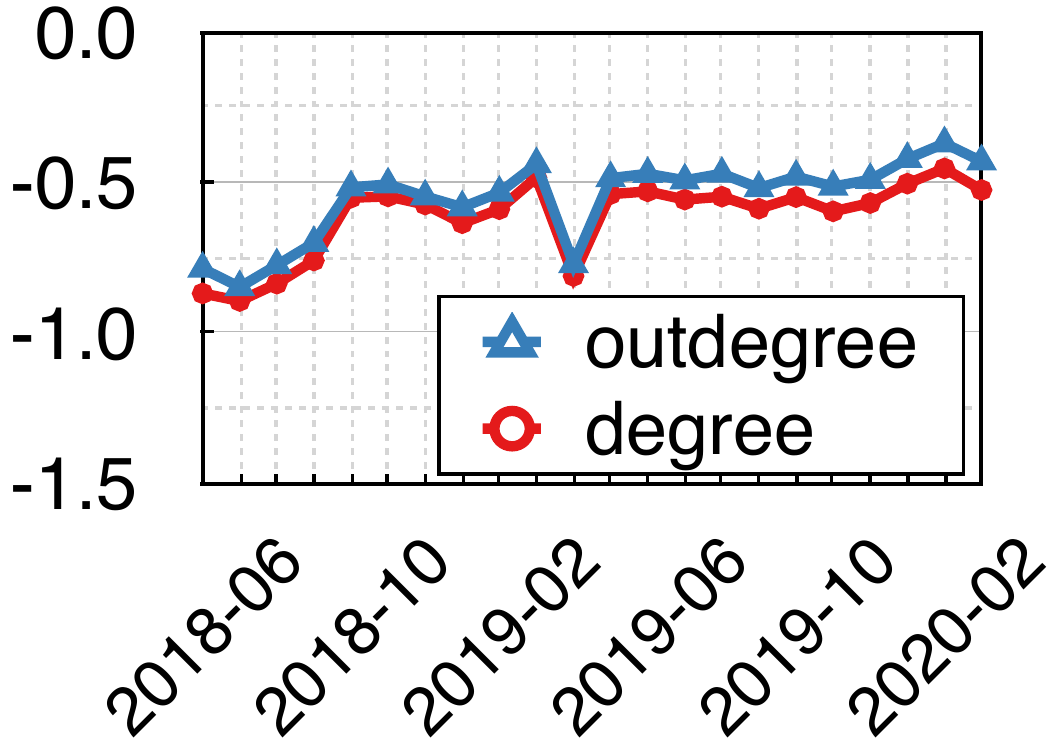}
         \caption{EOSIO}
         \label{fig:alpha-acg-eos}
     \end{subfigure}
        \caption{$\alpha$ of degree/outdegree distribution of blockchain platforms over time in terms of $ACG$.}
        \label{fig:alpha-acg}
\end{figure}

As each node can only be created once, this leads to a constant indegree value for each node. Hence, we only measure the distribution of outdegree and degree, and the evolution of the corresponding metrics $\alpha$, as shown in Fig.~\ref{fig:alpha-acg}.

For Ethereum, prior to mid-2017, the values of $\alpha$ show volatility, which we believe is caused by the relatively small number of $T_a$ during this period (see Fig.~\ref{fig:traces-acg}). From then on, both platforms exhibit an upwards trend for the $\alpha$ of the outdegree distribution, especially in EOSIO. In other words, the account creation behavior has become increasingly centralized. For example, account \texttt{genialwombat} which belongs to Wombat Wallet~\cite{wombat}, offers a free account creation service for its users. This wallet has created more than 32.2K accounts. Furthermore, the top 1\% of accounts in EOSIO have created 97.3\% of all accounts in Mar. 2020.
Interestingly, we can observe two sudden decreases in both sub-graphs: Aug. 2018 in Ethereum and Apr. 2019 in EOSIO. Combining the off-chain information we collected, we conclude that these two outliers were caused by an attack against a Fomo3D-like game called Last Winner~\cite{last-winner} and the EOS Global event, respectively. The former took advantage of lots of newly created accounts to attack the DApp. This account creation process followed a multi-level structure~\cite{last-winner} leading to invoking enormous $T_a$.

\textbf{\textit{Insight: }}
\textit{Except for the early stage of Ethereum, the $\alpha$ of the outdegree distributions show an upward trend for both platforms. The exceptions are two troughs, which resulted from attacks against Last Winner and EOS Global, respectively.
Especially in EOSIO, the upwards $\alpha$ indicates the emergence of large account creation providers which made the invocations of account creation transactions more centralized.}

\begin{figure}[t]
\centerline{\includegraphics[width=\columnwidth]{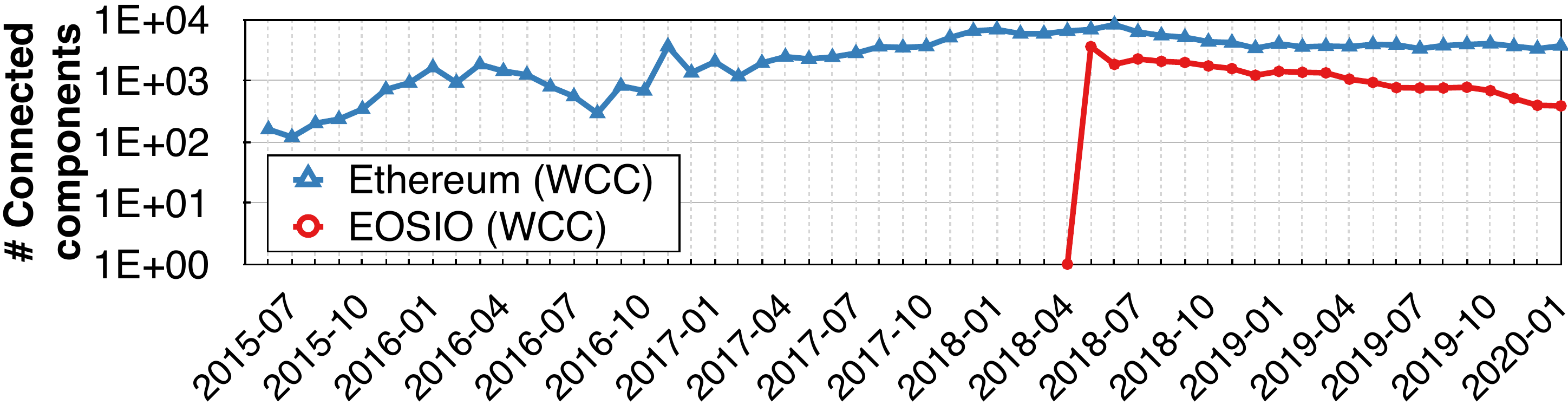}}
\caption{The number of WCC in different blockchain platforms over time in terms of $ACG$.}
\label{fig:cc-acg}
\end{figure}

\subsection{Connected Components}
\label{sec:blockchain-evolution:acg:cc}
Fig.~\ref{fig:cc-acg} shows the number of WCCs in Ethereum and EOSIO.
Ethereum shows an upward trend followed by a decline in terms of the number of WCC. 
In EOSIO, it was also gradually decreasing (due to the emergence of account creation services).
Moreover, we observe an obvious trough during Oct. 2016 in Ethereum, which had 294 WCCs.
We believe this is related to the DoS attack, which created a large number of accounts and prevented others from creating accounts properly.
Besides, the strange 1 WCC in EOSIO in Jun. 2018 was because the account creation tree composed of invoked $T_a$ has a common root node: official account \texttt{eosio}.

\textbf{\textit{Insight: }}
\textit{The account creation process has become more centralized, causing the number of WCC to go down continuously. This further reflects the tighter relationship between smart contracts in terms of account creation behavior.
}

\section{Evolution of Contract Invocation}
\label{sec:blockchain-evolution:cig}

\subsection{Graph Construction}
Similar to $ACG$, we do not discuss $T_c$ in Bitcoin.
We treat all the traces in Ethereum that are targeted to any smart contract as $T_c$.
As for EOSIO, we excluded the transfer actions to \texttt{eosio.token} and the account creation action to \texttt{eosio}. All the other actions are treated as $T_c$.
Consequently, we collect all the $T_c$ as $(v_i, v_j, w, t)$.
Identical to the process introduced in \S\ref{sec:blockchain-evolution:mtg:construction}, the $CIG$ is defined as a \textit{weighted directed graph}: $CIG = (V, E)$, where $E = \{(v_i, v_j, w^*, d) | v_i, v_j \in V, w^* \in \mathbb{R}^+\}$ and $V$ is the set of nodes.
In total, we have created 79 CIGs, 57 for Ethereum and 22 for EOSIO.

\subsection{Number of Traces}
\label{sec:blockchain-evolution:cig:traces}
\begin{figure}[t]
\centerline{\includegraphics[width=\columnwidth]{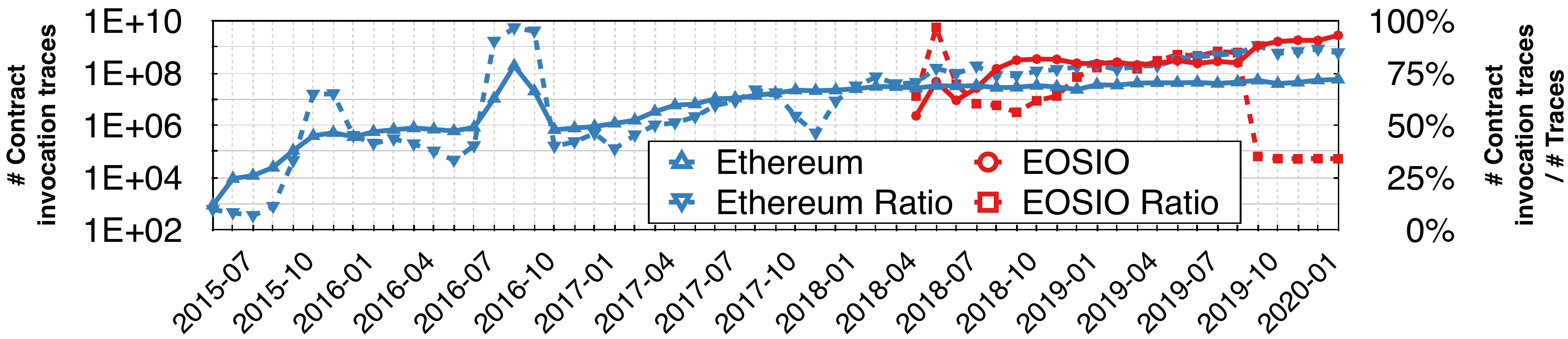}}
\caption{The evolution of contract invocation traces.}
\label{fig:traces-cig}
\end{figure}

Fig.~\ref{fig:traces-cig} depicts the number of $T_c$ and its corresponding ratio across time.
All these four lines fluctuate wildly.
For Ethereum, except for the spike in Oct. 2016 caused by the DoS attack which led to the number of $T_c$ exceeding 189 million, its absolute number shows a gradual upward trend.
Moreover, the growth of the ratio is dramatic: from the launch of Ethereum to Mar. 2020, the ratio grew from 9.60\% to 84.69\%, thereby starting to dominate Ethereum's network. 

As for EOSIO, its absolute number was also rising and was about an order of magnitude more than that of Ethereum, reaching almost two orders of magnitude after Nov. 2019. However, the trends between its ratio and absolute numbers are inconsistent. To be specific, we can see an obvious spike on Jul. 2018 that resulted from two accounts: \texttt{blocktwitter}
and \texttt{chaintwitter}, which imported more than 64.9M and 21.0M $T_c$. 
Then, during late-2018, a gambling DApp led to frequent invocations between accounts, thereby resulting in a trough during this period. Since then, the ratio was continuously rising till Nov. 2019, since when the EIDOS's mechanism almost fixes the ratio to 33.3\%.

\begin{figure}[t]
     \centering
     \begin{subfigure}[t]{0.8\columnwidth}
         \centering
         \includegraphics[width=\textwidth]{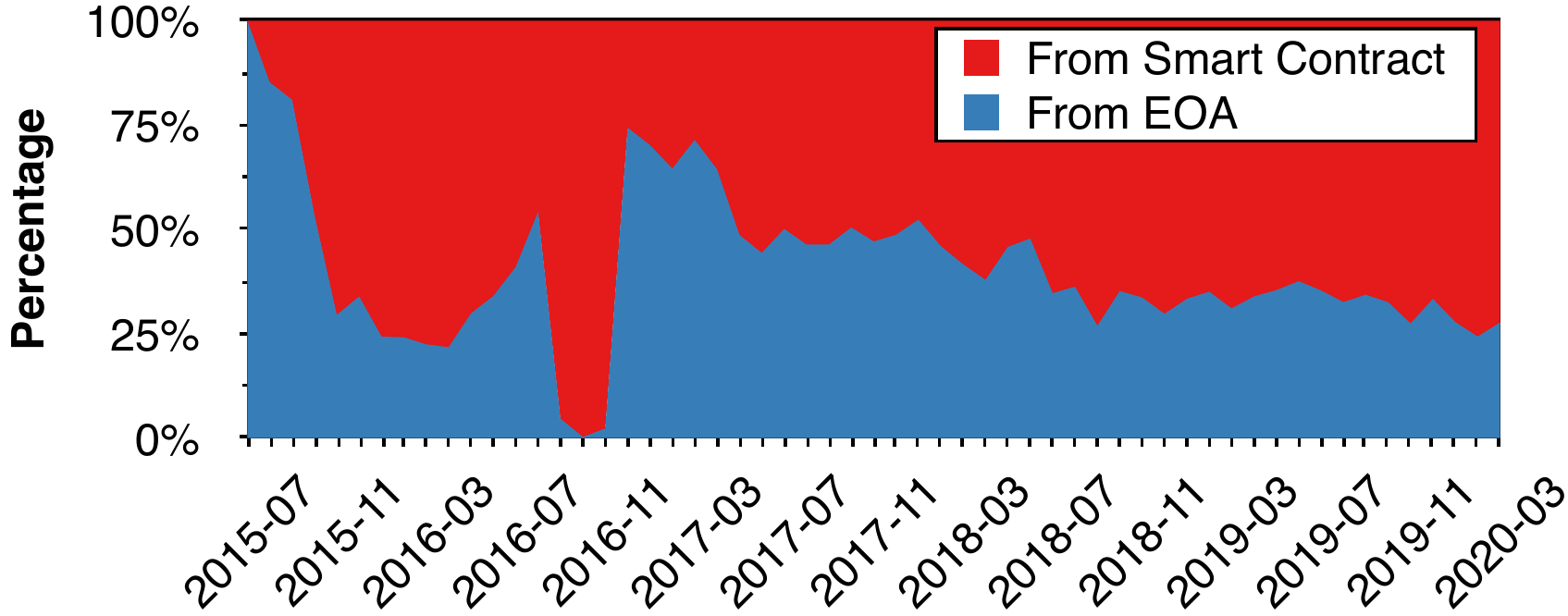}
         \caption{Ethereum}
         \label{fig:percentage-cig-eth}
     \end{subfigure}
     \begin{subfigure}[t]{0.8\columnwidth}
         \centering
         \includegraphics[width=\textwidth]{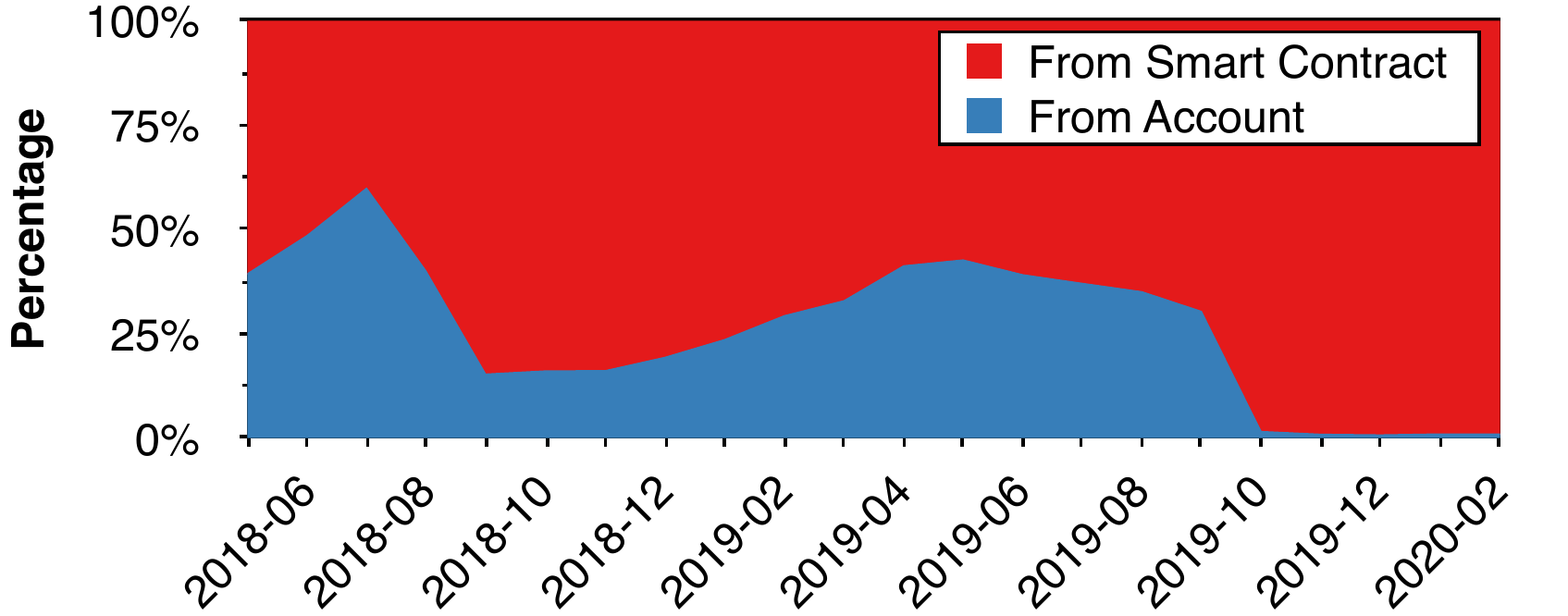}
         \caption{EOSIO}
         \label{fig:percentage-cig-eos}
     \end{subfigure}
        \caption{The distribution of initiators of $T_c$.}
        \label{fig:percentage-cig}
\end{figure}

The role of invokers who initiated $T_c$ is also interesting, as shown in Fig.~\ref{fig:percentage-cig}.
The DoS attack in Ethereum (Oct. 2016) is mainly initiated by smart contracts. 
Since then, smart contracts increasingly invoke other smart contracts.
But in EOSIO, a different tendency is observed: users in EOSIO preferred to invoke smart contracts from their own accounts. In May 2019, 41.42\% of $T_c$ was initiated by regular accounts. 
The reason is unclear, however, from Nov. 2019, EIDOS started to dominate the ecosystem, resulting in the percentage of smart-contract-invoking invocations rising to 98.27\%.

\textbf{\textit{Insight: }}
\textit{The number of smart contract invocations has been increasing across both platforms, confirming their growing use.
However, the number and the corresponding ratio were inconsistent in EOSIO, as they were significantly affected by gambling DApps and EIDOS.}

\subsection{Degree}
\label{sec:blockchain-evolution:cig:degree}
\begin{figure}[t]
     \centering
     \begin{subfigure}[t]{0.63\columnwidth}
         \centering
         \includegraphics[width=\textwidth]{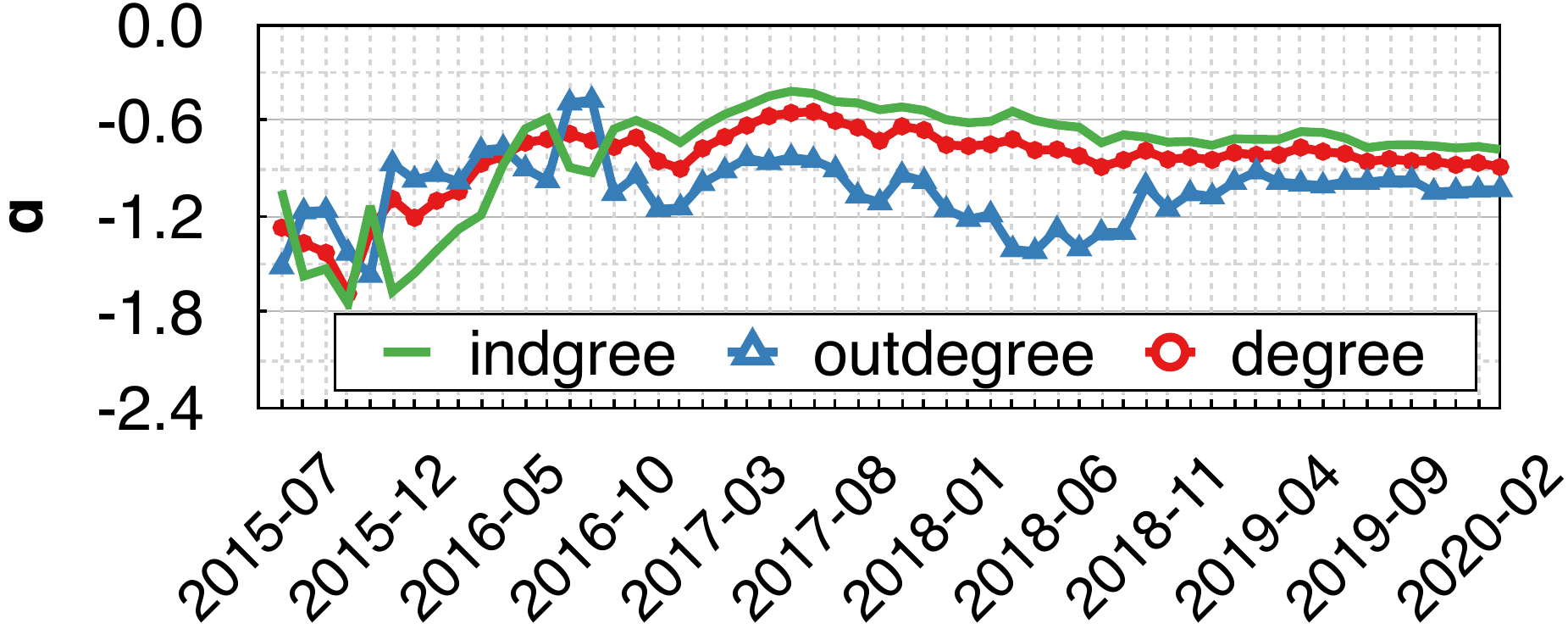}
         \caption{Ethereum}
         \label{fig:alpha-cig-eth}
     \end{subfigure}
     \begin{subfigure}[t]{0.36\columnwidth}
         \centering
         \includegraphics[width=\textwidth]{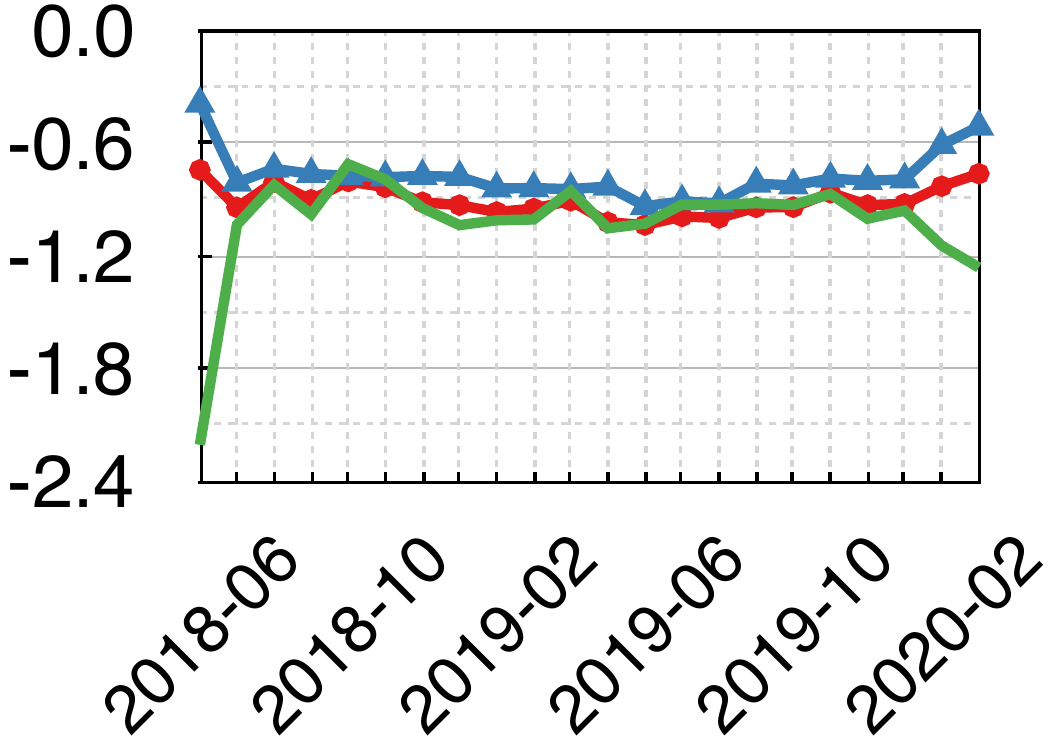}
         \caption{EOSIO}
         \label{fig:alpha-cig-eos}
     \end{subfigure}
        \caption{$\alpha$ of degree/indegree/outdegree distribution of blockchain platforms over time in terms of $CIG$.}
        \label{fig:alpha-cig}
\end{figure}

Fig.~\ref{fig:alpha-cig} illustrates the degree distribution.
Interestingly, the $\alpha$ of outdegree distribution of EOSIO is overall higher than that of Ethereum. The $\alpha$ of the indegree distribution, however, is on the opposite position. It indicates that, \textit{most accounts in Ethereum invoke fewer contracts than accounts in EOSIO, but there exist some \textit{super nodes} that accept most of the $T_c$ in Ethereum.} 
Moreover, in EOSIO, the $\alpha$ of the outdegree distribution is almost higher than the indegree's all the time. This illustrates the existence of some killer DApps that invokes many $T_c$.
The significant upward trend in 2020 in Fig.~\ref{fig:alpha-cig-eos} reflects that the initiation of $T_c$ became more centralized. Recall the mechanism of EIDOS we briefly introduced in \S\ref{sec:blockchain-evolution:mtg:traces}, which initiates lots of EIDOS token transfers to benefit users. Alone, it accounted for 92.2\% of $T_c$ in Nov. 2019, which indicates the tendency towards centralization in EOSIO.
Additionally, in Ethereum, for the $\alpha$ of the outdegree distribution, the trough during the whole 2018 in Fig.~\ref{fig:alpha-cig-eth} is highly conspicuous. The reason is the emergence of some popular DeFi DApps who attract users to invoke them; this finally makes the distribution of outdegree less centralized.

\textbf{\textit{Insight: }}
\textit{The $\alpha$ of the outdegree/indegree distributions in EOSIO and Ethereum follow opposite trends. This reflects the centralization of the outdegree in EOSIO, caused by popular gambling DApps and mis-behaviors; and the centralization of the indegree in Ethereum due to the emergence of DeFi DApps.
}

\subsection{Pearson Coefficient ($R$)}
\label{sec:blockchain-evolution:cig:pearson}
\begin{figure}[t]
\centerline{\includegraphics[width=\columnwidth]{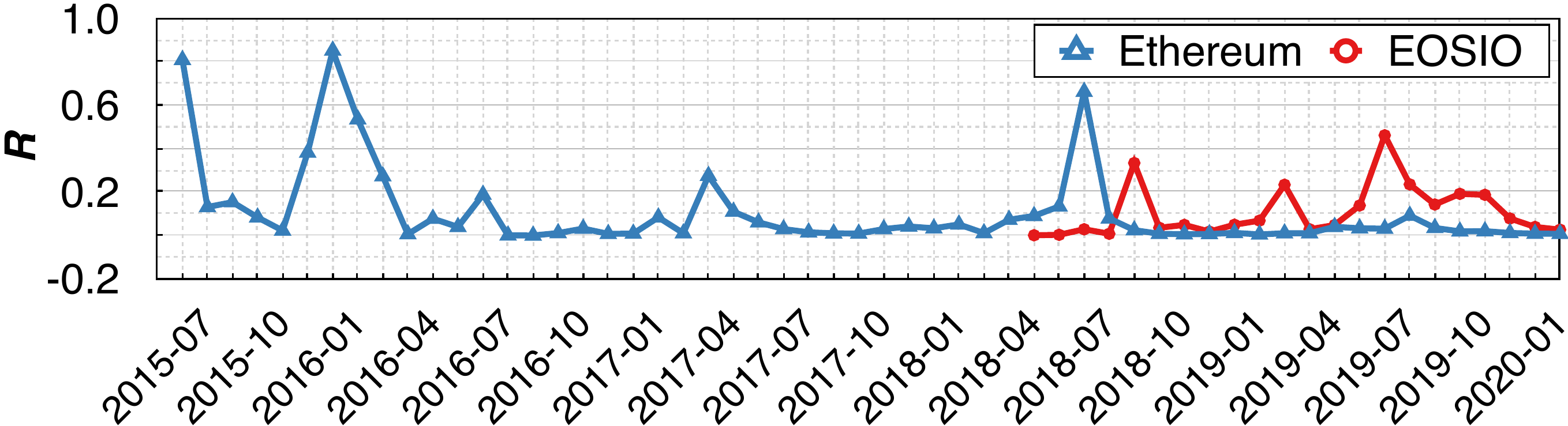}}
\caption{Pearson's correlation between indegree and outdegree of blockchain platforms over time in $CIG$.}
\label{fig:pearson-cig}
\end{figure}

Fig.~\ref{fig:pearson-cig} gives the $R$ between indegree and outdegree in terms of contract invocation.
We see there was little correlation between the indegree and outdegree in the $CIG_{eth}$ and $CIG_{eos}$, and there never existed a negative correlation.
That said, in some months, the value of $R$ rises sharply. This is caused by popular DApps or malicious behaviors; these have such a large indegree and outdegree that they are able to affect the entire network.
For example, in Aug. 2018, a gambling game called \textit{Last Winner} launched in Ethereum; and in Aug. 2019, a popular DApp \textit{pornhashbaby} appeared in EOSIO. The former resulted in 876K indegree and 311K outdegree, accounting for more than 45\% gambling related $T_c$ in $CIG_{eth}^{2018-08}$.

\textbf{\textit{Insight: }}
\textit{The indegree and outdegree of nodes in both platforms had little correlation in terms of contract invocation. However, this metric can be severely impacted by misbehaviors or popular DApps.
}

\subsection{Connected Components}
\label{sec:blockchain-evolution:cig:cc}
\begin{figure}[h]
\centerline{\includegraphics[width=\columnwidth]{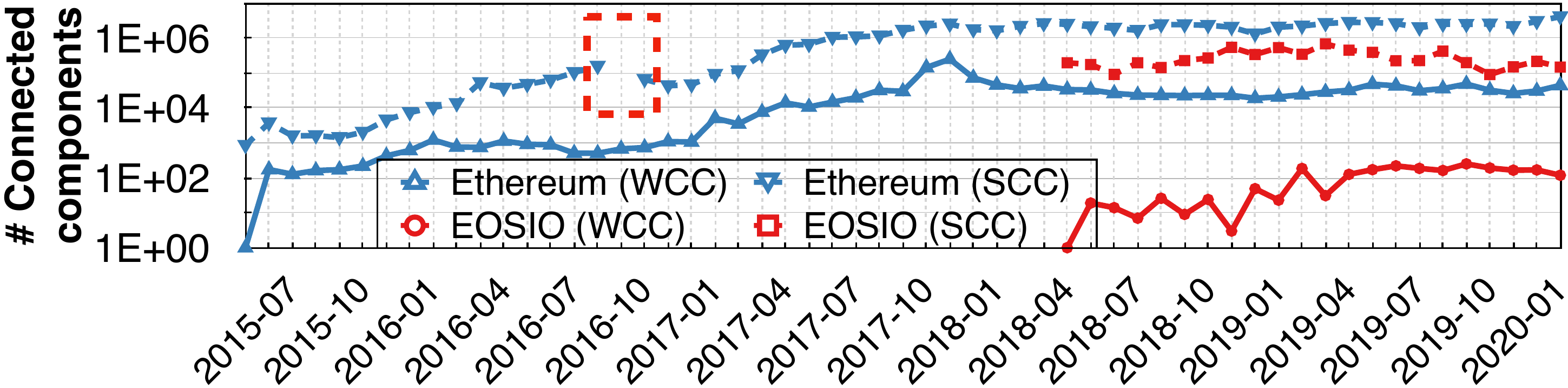}}
\caption{The number of WCC and SCC in Ethereum and EOSIO over time in terms of $CIG$}
\label{fig:cc-cig}
\end{figure}

Fig.~\ref{fig:cc-cig} shows the number of WCCs and SCCs in Ethereum and EOSIO in $CIG$s.
These have been gradually increasing. 
The peak on Jan. 2018 in Ethereum's WCC is obvious, caused by the sudden decrease of users in Bittrex due to a hack event~\cite{bittrex-hack}.
Moreover, the number of WCCs in EOSIO is nearly two orders of magnitude smaller than Ethereum's.
Combining Fig.~\ref{fig:traces-cig} and Fig.~\ref{fig:alpha-cig-eos}, we conclude that \textit{users on EOSIO were calling contracts more frequently}.

Besides, some accounts initiated a large number of $T_c$, 
which covered numerous contracts and kept the number of WCC at a low level. The maximum WCC has covered at least 99.24\% of nodes in the corresponding months. However, for SCC, we can conclude that the callback from the smart contract is unusual. This may be due to two reasons: 1) the initiating account is a regular account, thus it is unable to callback by a $T_c$; and 2) the callback trace just sends some native tokens, which is classified as a $T_m$. 
Note that there is a missing point of \#SCC in Nov. 2016. This is because Ethereum officially deployed a sweeper contract\footnote{0xa43ebd8939d8328f5858119a3fb65f65c864c6dd} that initiated more than 16 million one-way $T_c$ to remove all the redundant and meaningless contracts created by the DoS attack in October. This resulted in a large number of SCCs ($>$16M).

\textbf{\textit{Insight: }}
\textit{
The growing number of accounts, indicate smart contracts are gaining increasing popularity.
}

\section{Characterizing Outliers}
\label{sec:abnormal-behaviors}

\subsection{Definition of Outlier}

Our previous exploration suggests that some events (e.g., DoS attacks) and influential accounts have a large impact on ecosystems. 
In this section, we explore such events and addresses. \textit{Note our outlier analysis is not intended to identify attacks, as many attack events may not cause noticeable changes to the overall ecosystems.} Instead, our purpose is to pinpoint what kinds of events/DApps can facilitate or impede the development of the blockchain ecosystems.
For all metrics considered, some data points deviate significantly from the average~\cite{outlier}. We refer to these points as \textit{outliers}.

\subsection{Detecting and Labelling Outliers}
\label{sec:abnormal-behaviors:method}

\subsubsection{Detecting Outliers.} 
We use a simple but efficient algorithm, \textit{z-score}~\cite{z-score}, to flag outliers in each time series. Z-scores quantify the unusualness of an observation. Z-scores are the number of standard deviations above and below the mean that each value falls. Specifically, if the value of a metric for a month is $x$, the z-score would be: $z=\frac{x-\mu}{\sigma}$, where $\mu$ is the average of $x$ and corresponding values before and after three months (7 values in total), and $\sigma$ is the standard deviation for these 7 values.
Therefore, the absolute value of the z-score represents the distance between the $x$ and its adjacent elements mean in units standard deviation.

\subsubsection{Labelling Outlier}
\label{sec:abnormal-behaviors:detecting:labelling}
We propose a semi-automated method to label outliers.
Our idea is that \textit{outliers result from suddenly popular DApps or misbehaviors}. Thus, removing the responsible node(s) can make the absolute value of z-score smaller than a threshold.
For each metric, we have different strategies to identify the responsible node(s).

Specifically, (1) for the number of trace metric, the node with the highest degree (named the \textit{supernode}) in the corresponding graph will be identified; 
(2) for $\alpha$ of the degree distribution, there are two cases: trough or peak, corresponding to dispersion and centralization, respectively. Therefore, for the former, the strategy tries to identify the node(s) who import many nodes with a small degree; for the latter, the supernode with the highest degree/indegree/outdegree will be labeled;
(3) based on our extensive manual investigation, a node with high indegree and outdegree simultaneously is likely to push the $R$ higher. Therefore, the strategy focuses on the node with the highest product of indegree and outdegree;
(4) the significant change in the number of connected components is always related to the (dis)appearance of nodes that are directly connected to lots of nodes in the network. To this end, the nodes with the most one-way or two-way edges with the others will be identified.

After the node(s) are identified according to the above strategy, it and its connected edges in the corresponding graph will be removed. We then recalculate the corresponding metric to see if the absolute value of z-score drops lower than a threshold. If not, the above process is repeated until the outlier disappears.
To guarantee accuracy, we manually recheck all removed nodes to confirm they are the actual responsible ones.
In other words, for each outlier, \textit{our method can extract a set of responsible nodes that are related to attack/spam events or killer DApps.}

We classify the outliers into two major groups: \textit{Killer DApps} and \textit{Misbehaviors}, as shown in Table~\ref{table:category-outliers}.
The Killer DApp category indicates the outlier is due to the activity of a popular DApp. 
The misbehavior category is divided into three sub-categories: \textit{Attack}, \textit{Resource Manipulation}, and \textit{Spam}. An attack means the outlier has been caused by a well-known attack, like the DAO reentrancy attack~\cite{the-dao}. Resource manipulation indicates the outlier is caused by an exploitation on the resource model, like EIDOS (see \S\ref{sec:background:general:consensus}). The spam sub-category covers spam advertisements and bot accounts. 
Note that the z-score algorithm also flags some outliers in the infant stage of the blockchains. We omit these, as a few transactions can lead to a volatile swing on the value of a metric.

\subsection{Results}
\label{sec:abnormal-behaviors:overall}

\begin{table*}[t]
\caption{The overall result of outlier labelling.}
\centering
\resizebox{0.7\textwidth}{!}{%
\begin{tabular}{l|ccccccc|ccc|c}
\toprule	
                  & \multicolumn{7}{c|}{\textbf{Killer DApp}}                                         & \multicolumn{3}{c|}{\textbf{Misbehavior}}                                                                 & \multirow{2}{*}{\textbf{Total}} \\ \cline{2-11}
                  & DeFi       & Exchange   & Gambling   & Game       & Platform   & Token      & Tool       & Attack     & \begin{tabular}[c]{@{}c@{}}Resource\\Manipulation\end{tabular} & Spam        &                                 \\ \midrule
\textbf{Bitcoin}  & 0          & 0          & 2          & 2          & 5          & 0          & 0          & 3          & 0                                                                              & 1           & \textbf{13}                      \\ \midrule
\textbf{Ethereum} & 2          & 10          & 3          & 6          & 0          & 8          & 2          & 2          & 13                                                                             & 3           & \textbf{49}                     \\ \midrule
\textbf{EOSIO}    & 0          & 0          & 5          & 0          & 1          & 1          & 1          & 0          & 2                                                                              & 15          & \textbf{25}                     \\ \midrule
\textbf{Total}    & \textbf{2} & \textbf{10} & \textbf{10} & \textbf{8} & \textbf{6} & \textbf{9} & \textbf{3} & \textbf{5} & \textbf{15}                                                                    & \textbf{19} & \textbf{87}                    \\
\bottomrule
\end{tabular}
}
\label{table:category-outliers}
\end{table*}

Table~\ref{table:category-outliers} shows the overall results (see Table~\ref{table:anomalies-1} to Table~\ref{table:anomalies-3} in Appendix for details), 87 outliers are identified in total: 13 for Bitcoin, 49 for Ethereum, and 25 for EOSIO. 
As concluded before, Bitcoin is the most stable platform. Though EOSIO fluctuates in several plots, due to its relative short history, its number of outliers is below Ethereum's.

Among the 87 outliers, 45\% (39) belongs to the misbehavior category. Moreover, most of them are in the resource manipulation and spam sub-categories. Ethereum suffers more from malicious exploitation on resources. The main reason is the DoS attack, which was so powerful that it significantly altered many metrics in Oct. 2016. Due to the resource model in EOSIO (see \S\ref{sec:background:general:consensus}), it is more likely to suffer from spam.
This is partly due to the EOS Global event, which created many useless accounts and transferred EOS between them. The spam advertisements in EOSIO also played an important role (detailed in \S\ref{sec:abnormal-behaviors:case-study:spam}).
Moreover, we observe that, compared to EOSIO, Ethereum is impacted more by killer DApps. These apps are evenly distributed across categories. This likely indicates that Ethereum is still preferred by DApp developers. The only killer app in EOSIO is the pornhashbaby, which brought significant traffic until it was shutdown.

\subsection{Case Studies}
\label{sec:abnormal-behaviors:case-study}
We next briefly introduce two case studies, while the details can be found in Appendix \S\ref{sec:appendix:case-study}.

\subsubsection{EIDOS Event}
\label{sec:abnormal-behaviors:case-study:EIDOS}
From Fig.~\ref{fig:traces} we observe that, in Nov. 2019, the number of traces in EOSIO jumped by 1,009.82\% compared to the preceding month (reaching 3.21 billion pieces). Surprisingly, over 99\% of the transactions are related to the EIDOS contract. However, only a few hundred accounts interacted with the contract. This suggests that billions of transactions were introduced by a small number of addresses.
We further calculate the total volume of money transferred to the EIDOS contract, and the average EOS per $T_m$. After a slightly increase in Dec. 2019, both of these two metrics dropped significantly. In Mar. 2020, only 1.43 million EOS were transferred to EIDOS (around 6.53\% of the peak). Meanwhile, the average EOS per $T_m$ decreased to $6 \times 10^{-4}$, while $1 \times 10^{-4}$ is the minimum allowed transfer amount in EOSIO. We conclude that players were becoming rational and centralized, and a number of services have emerged~\cite{eidos-miner-1, eidos-miner-2} that allow participants to maximize benefits at minimal cost.
EIDOS also overtook other DApps, e.g., previously popular gambling DApps, and has had a significant negative impact on the entire ecosystem.

\subsubsection{Spam Advertisement}
\label{sec:abnormal-behaviors:case-study:spam}
During the analysis of Fig.~\ref{fig:alpha-mtg-eos}, we have observed two strange spikes in the $\alpha$ of the outdegree distribution (Mar. and Oct. 2019).
We discover that these two outliers are due to \textit{spam advertisement}.
Spammers have taken advantage of the resource model in EOSIO, which allows users to initialize spam advertisements to cover huge amounts of accounts; specifically, in the memo field~\cite{eos-memo} of transactions, users can write use this to free-text. We see that spammer write things like bait-and-switch advertisements.
Using $MTG$ and $CIG$, we further propose an automated method to identify spammers (see Appendix \S\ref{sec:appendix:case-study:spam} for details). The key idea is that within a given time frame, the spammer would initiate spam advertisements carrying identical spam content by the lowest cost to cover as many accounts as possible. Thus, we can group together small transactions with identical content to flag spam candidates.
Through this, we identify \textbf{206} distinct spam accounts in EOSIO, and we identify two account family: \texttt{peostoken} and \texttt{defind.io}, who are responsible for the spikes located in Mar. 2019 and Oct. 2019.

\section{Discussion}

\subsection{Cross-chain Comparison}

\noindent
\textbf{Bitcoin.} Bitcoin is a value-transfer network, and does not support features like smart contracts. Moreover, its simple but effective transaction fee mechanism guarantees that it is hard to be abused. To this end, Bitcoin has been in a steady growth and has been hard to be manipulate according to its number of transfer transactions and degree distribution.

\noindent
\textbf{Ethereum.} Ethereum is still evolving, especially in terms of contract invocation. Prior to 2018, the most dominant transaction in Ethereum is the Ether transfer. The price of Ether would always incentivize a more frequent and centralized Ether transfer between users and exchanges. Since then, the emergence of DeFi has increased the number of interactions between smart contracts. Along with the growing number of transactions, the relatively stable ratio of newly created smart contracts also implies the liveness in Ethereum.

\noindent
\textbf{EOSIO.}
Via DPoS, EOSIO is able to carry out hundreds of times the traces than Bitcoin and Ethereum (who adopt PoW). 
Prior to Nov. 2019, though there was a slight decline in the number of transfers and contract invocations, the popularity of gambling DApps has attracted many users. However, the appearance of EIDOS almost collapsed the entire network: The TPS was pushed to its limit, and all other normal transactions suffered high congestion.

\subsection{Threats to Validity and Future Work}
This study carries several limitations.
First, we primarily study the evolution of the ecosystems based on graph analysis and, by definition, the metrics used are limited. Although we have considered the most widely used metrics, more metrics (e.g., global clustering coefficient~\cite{clustering}, page rank~\cite{page-rank}, reciprocity~\cite{reciprocity}) could be adopted to gain further understanding of ecosystems.
Second, in several cases, we have relied on heuristics and manual validation due to the paucity of ground-truth data. For example, during the outlier analysis, we have manually confirmed the underlying reasons for several cases. We agree that automated approaches (e.g., via machine learning) could be used to facilitate the process.
Third, our study relies solely on transaction analysis. We believe additional code analysis (of the smart contracts) could help us to gain a better understanding of ecosystems.
Finally, although some highly influential attacks and unreported scams (e.g., the scam advertisements in EOSIO) have been discovered by us, we do not consider a number of known attacks in the community (as they did not have a noticeable impact on the overall ecosystem). 
For future work, we plan to address the above limitations.

\section{Related Work}
\label{sec:related}

\textbf{Blockchain Transaction Analysis.}
\label{sec:related:tx-analysis}
Some previous works have focused on blockchain transaction analysis for a single blockchain platform, e.g., Bitcoin~\cite{awan2017blockchain,ober2013structure, ron2013quantitative, houstudy}, Ethereum~\cite{chen2018understanding, wu2019t, chen2020traveling, kiffer2018analyzing, lee2020measurements, zhao2021temporal}, EOSIO~\cite{huang2020understanding, perez2020revisiting} and Monero~\cite{moser2018empirical}.
Ober et al.~\cite{ober2013structure} and Awan et al.~\cite{awan2017blockchain} both analyzed the topology and dynamics of the Bitcoin transaction graph.
Chen et al.~\cite{chen2018understanding} used graph analysis to explore the characteristics of transactions in Ethereum.
Similarly, Chen et al.~\cite{chen2020traveling} and Huang et al.~\cite{huang2020understanding} also adopted graph analysis, but they focused on the behaviors of ERC20 tokens and fraudulent activities in EOSIO, respectively.

\noindent
\textbf{Smart Contract \& DApp Analysis.}
\label{sec:related:sc-analysis}
There are some works \cite{wu2019empirical, wang2020identifying} dedicated to exploring the ecosystem of DApps.
For example, Wu et al.~\cite{wu2019empirical} defined the popularity of DApps by some metrics across different categories. They measured the reuse of DApps, and the transformation of a traditional application to a DApp. Wang et al.~\cite{wang2020identifying} classify DApps and users behaviors according to the patterns extracted from transactions.
Moreover, some works~\cite{suevil, chen2020survey, he2020characterizing, ji2020deposafe, he2021eosafe} have focused on the security issue of DApps across platforms.
For example, Chen et al.~\cite{chen2020survey} conducted a detailed survey about vulnerabilities and defenses related to smart contracts in DApps. Su et al.~\cite{suevil} and He et al.~\cite{he2020characterizing} measured the attack instances and clone behaviors against DApps, respectively.

\section{Conclusion}

This paper has presented a longitudinal comparative study of three representative blockchains: Bitcoin, Ethereum and EOSIO. 
We utilized billions of transaction records to offer a unique insight into the overall ecosystems.
These have revealed complementary trends amongst the three platforms, highlighting their differing evolutions. 
Although our findings have revealed some promising trends for these technologies, a number of ``outliers'' have also been identified and analyzed, e.g., abuse by spammers.
We believe that our research efforts contribute positively to inform best operational practices and will benefit the wider research community and regulators.


\bibliographystyle{ACM-Reference-Format}
\bibliography{sample-base}

\appendix
\newpage
\section{Money Transfer in Bitcoin, Ethereum and EOSIO}
\label{sec:appendix:money-transfer}
\begin{figure}[h]
     \centering
     \begin{subfigure}[t]{0.9\columnwidth}
		\includegraphics[width=\textwidth]{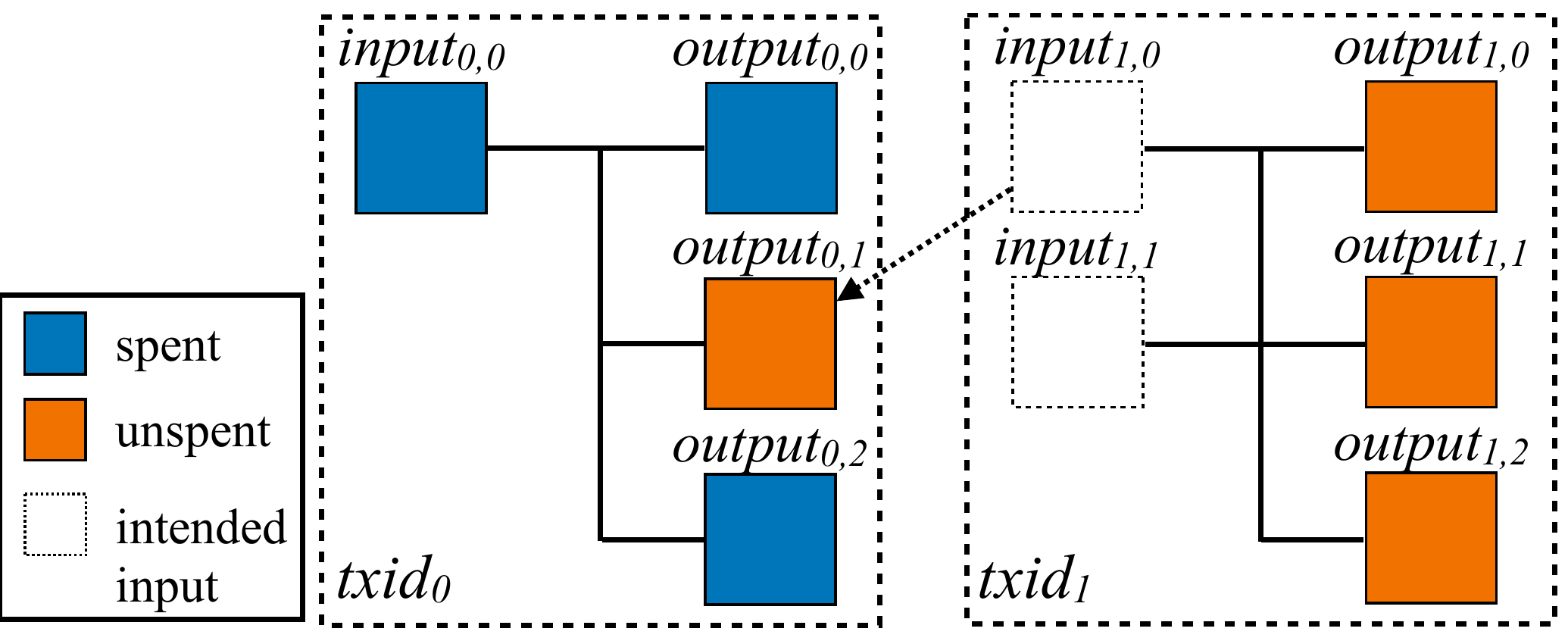}
		\caption{UTXO model of Bitcoin.}
		\label{fig:utxo}
	\end{subfigure}
     \begin{subfigure}[t]{0.9\columnwidth}
		\includegraphics[width=\textwidth]{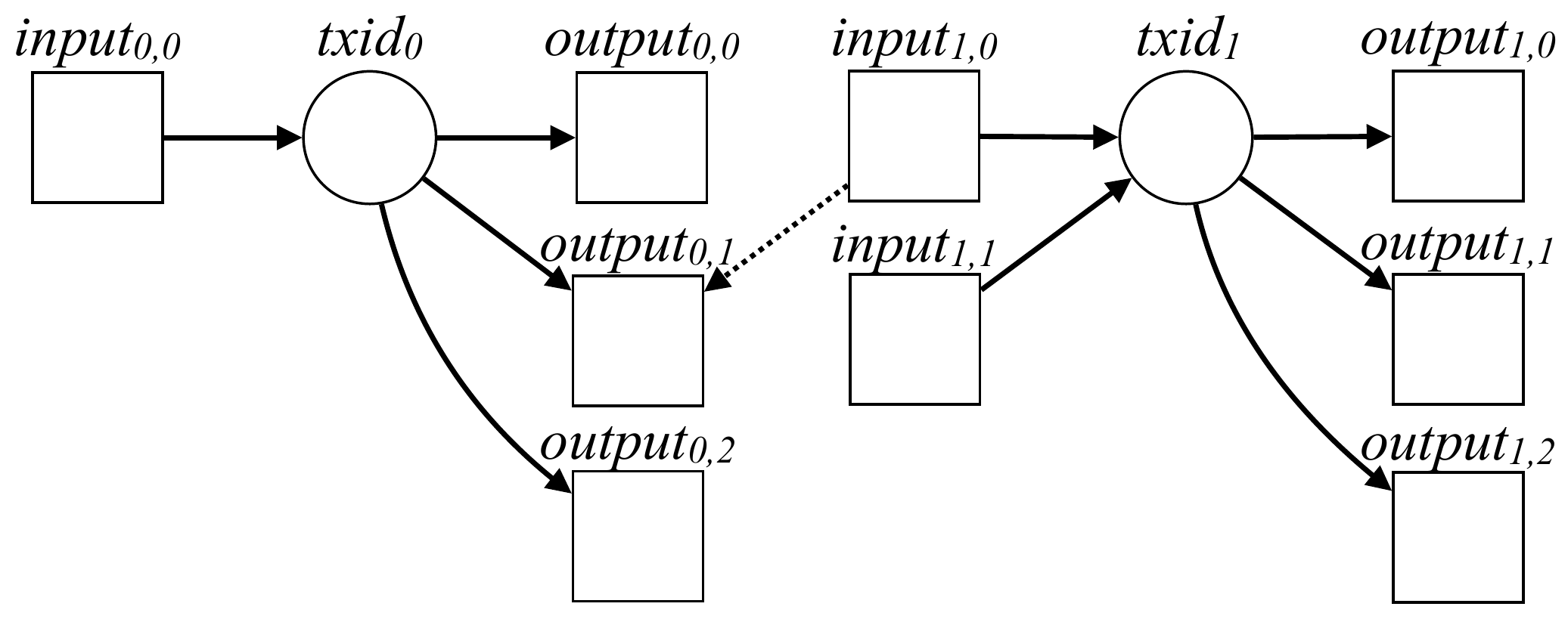}
		\caption{Its Structure of Money Transfer Graph.}
		\label{fig:transfer-btc}
	\end{subfigure}
        \caption{UTXO model and its corresponding structure of Money Transfer Graph.}
        \label{fig:bitcoin}
\end{figure}

Fig.~\ref{fig:utxo} shows the UTXO model adopted by Bitcoin, and Fig.~\ref{fig:transfer-btc} depicts the corresponding topology of $MTG_{btc}$.

\begin{figure}[h]
     \centering
     \begin{subfigure}[t]{0.9\columnwidth}
         \centering
         \includegraphics[width=\textwidth]{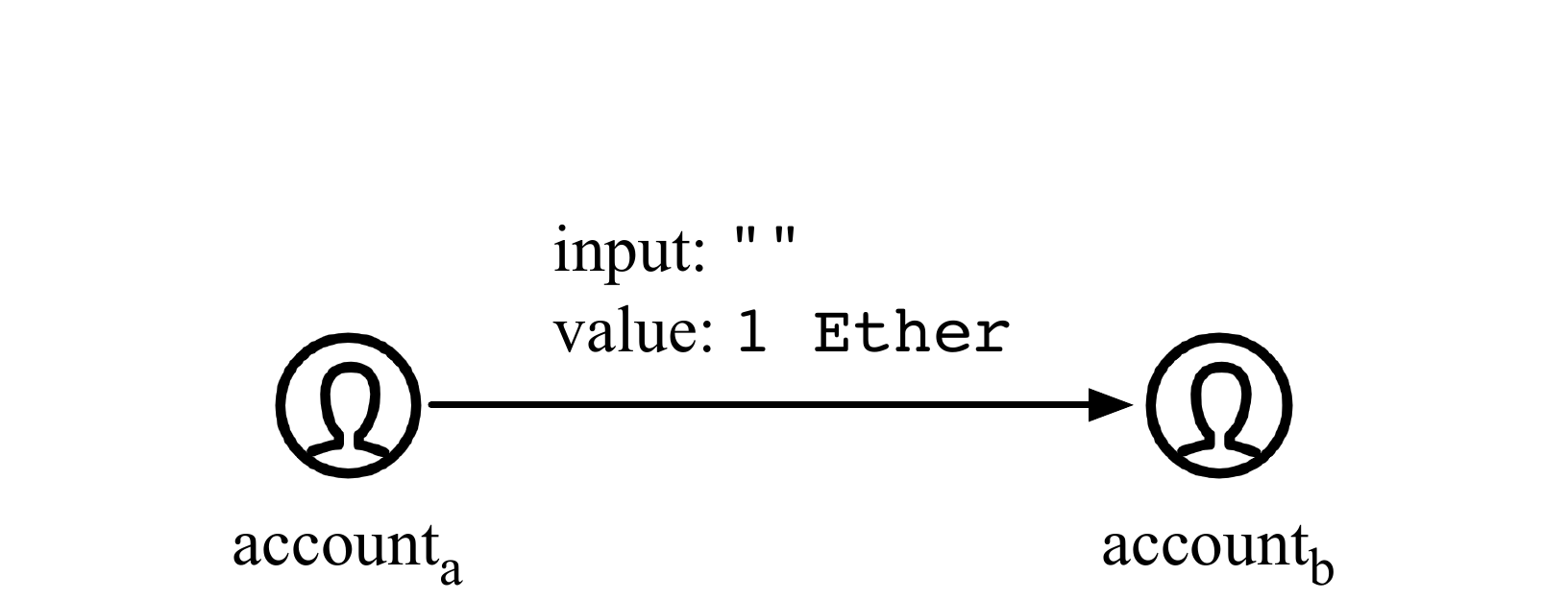}
         \caption{Transferring Ether in Ethereum}
         \label{fig:transfer-eth}
     \end{subfigure}
     \begin{subfigure}[t]{0.9\columnwidth}
         \centering
         \includegraphics[width=\textwidth]{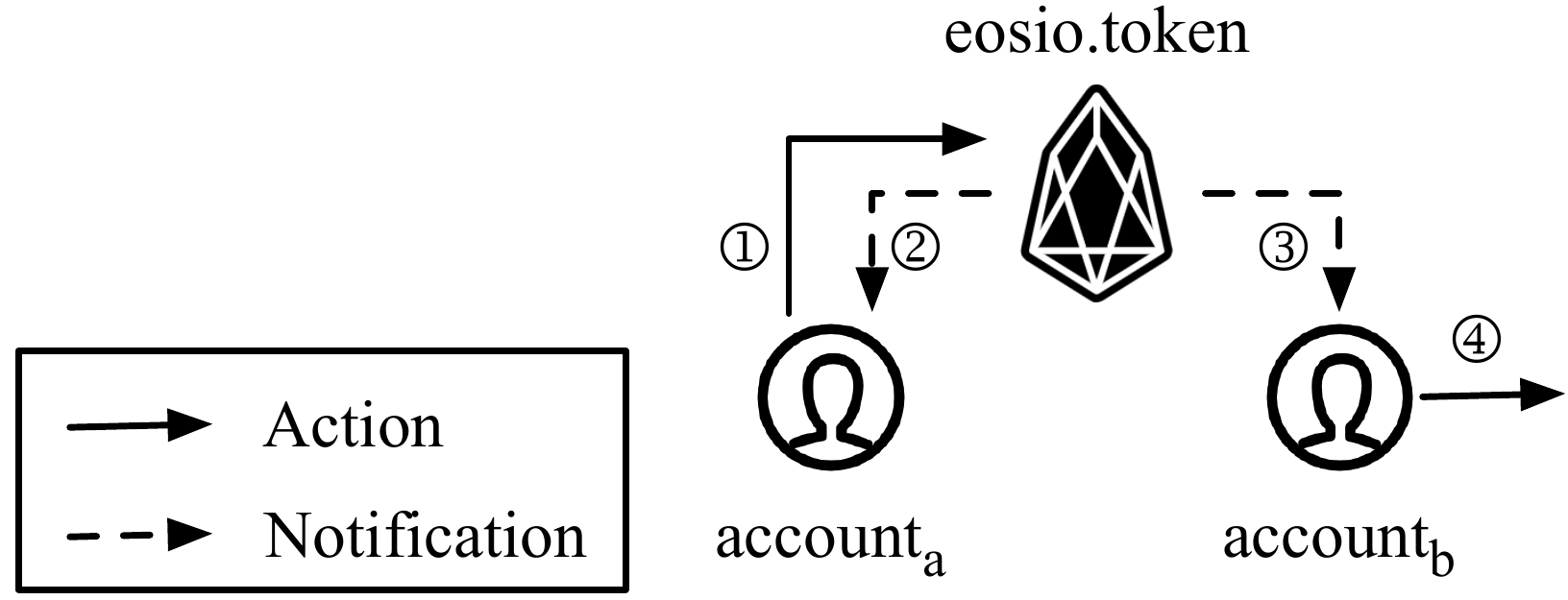}
         \caption{Transferring EOS in EOSIO}
         \label{fig:transfer-eos}
     \end{subfigure}
        \caption{Transferring official tokens.}
        \label{fig:transfer-token}
\end{figure}

Fig.~\ref{fig:transfer-eth} and Fig.~\ref{fig:transfer-eos} respectively illustrate the transferring Ether and EOS in Ethereum and EOSIO.

\newpage
\section{Outliers}
\label{sec:appendix:anomalies}
The detected outliers and their corresponding sub-category (category is omitted, details in Table~\ref{table:category-outliers}), as well as the reason, are shown in Table~\ref{table:anomalies-1} to Table~\ref{table:anomalies-3}.

\begin{table*}[h]
\caption{The detected outliers and its corresponding sub-category, as well as the specific reason.}
\centering
\resizebox{0.68\textwidth}{!}{%
\begin{tabular}{lccccc}
\toprule
\textbf{Chain}                 & \textbf{Date} & \textbf{Metric}     & \textbf{Figure}              & \textbf{Sub-category}                                                          & \textbf{DApp / Event}                      \\
\midrule
\multirow{8}{*}{\textbf{BTC}}  & 2012-02       & indegree $\alpha$    & Fig.~\ref{fig:alpha-mtg-btc}            & Platform                                                                       & Deepbit                                    \\
                               & 2015-07       & outdegree $\alpha$   & Fig.~\ref{fig:alpha-mtg-btc}            & Game                                                                           & 1Luc*        \\
                               & 2015-12       & \# WCC              & Fig.~\ref{fig:cc-mtg}                   & Game                                                                           & 1Luc*         \\
                               & 2012-02      & degree $\alpha$     & Fig.~\ref{fig:alpha-mtg-btc}                & Platform                                                                      & Deepbit        \\
                               & 2012-02      & outdegree $\alpha$  & Fig.~\ref{fig:alpha-mtg-btc}                & Platform                                                                      & Deepbit        \\
                               & 2010-11      & \# traces          & Fig.~\ref{fig:traces-mtg}                 & Spam                                                                          & 14mU** \\
                               & 2013-07       & \# traces          & Fig.~\ref{fig:traces}                     & Gambling                                                                      & Satoshi Dice
                               \\
                               & 2015-07       & \# traces         & Fig.~\ref{fig:traces}                     & Gambling                                                                      & LuckyBit hot wallet
                               \\
                               & 2012-01       & \# traces         & Fig.~\ref{fig:traces}                     & Platform                                                                      & Deepbit
                               \\
                               & 2012-02        & \# traces        & Fig.~\ref{fig:traces}              & Platform                                                                              & Deepbit   \\
                               & 2015-07        & \# traces         & Fig.~\ref{fig:traces}             & Attack                                                                            & Spam Attack  \\
                               & 2015-07        & \# traces         & Fig.~\ref{fig:traces-mtg}         & Attack                                                                            & Spam Attack \\
                               & 2015-07        & \# traces         & Fig.~\ref{fig:traces}             & Attack                                                                            & Spam Attack \\
\midrule
\multirow{30}{*}{\textbf{EOS}} & 2018-10       &$R$                  & Fig.~\ref{fig:pearson-cig}              & Gambling                                                                       & EOSTiger                                   \\
                               & 2018-10       &$R$                  & Fig.~\ref{fig:pearson-cig}              & Gambling                                                                       & Dice                                       \\
                               & 2018-10       &$R$                  & Fig.~\ref{fig:pearson-cig}              & Gambling                                                                       & eos sicbo                                  \\
                               & 2018-10       &$R$                  & Fig.~\ref{fig:pearson-cig}              & Gambling                                                                       & EOS.Win                                    \\
                               & 2019-04       &$R$                  & Fig.~\ref{fig:pearson-cig}              & Spam                                                                           & EOS Global                                 \\
                               & 2019-08       &$R$                  & Fig.~\ref{fig:pearson-cig}              & Platform                                                                       & pornhashbaby                               \\
                               & 2019-04       & indegree $\alpha$    & Fig.~\ref{fig:alpha-cig-eos}            & Spam                                                                           & EOS Global                                 \\
                               & 2019-04       &$R$                  & Fig.~\ref{fig:pearson-mtg}              & Spam                                                                           & EOS Global                                 \\
                               & 2019-11       &$R$                  & Fig.~\ref{fig:pearson-mtg}              & \begin{tabular}[c]{@{}c@{}}Malicious\\ Exploitation\\ on Resource\end{tabular} & EIDOS                                      \\
                               & 2019-04       & \# WCC              & Fig.~\ref{fig:cc-mtg}                   & Spam                                                                           & EOS Global                                 \\
                               & 2019-10       & outdegree $\alpha$   & Fig.~\ref{fig:alpha-mtg-eos}            & Spam                                                                           & defind.io                                  \\
                               & 2019-04       & \# traces           & Fig.~\ref{fig:traces-acg}               & Spam                                                                           & EOS Global                                 \\
                               & 2019-04       & trace ratio         & Fig.~\ref{fig:traces-acg}               & Spam                                                                           & EOS Global                                 \\
                               & 2019-04       & From Account        & Fig.~\ref{fig:percentage-acg-eos}       & Spam                                                                           & EOS Global                                 \\
                               & 2019-04       & degree $\alpha$      & Fig.~\ref{fig:alpha-acg-eos}            & Spam                                                                           & EOS Global                                 \\
                               & 2019-04       & outdegree $\alpha$   & Fig.~\ref{fig:alpha-acg-eos}            & Spam                                                                           & EOS Global                                 \\
                               & 2019-04       & \# SCC              & Fig.~\ref{fig:cc-acg}                   & Spam                                                                           & EOS Global                                 \\
                               & 2019-11       & degree $\alpha$     & Fig.~\ref{fig:alpha-cig-eos}                & \begin{tabular}[c]{@{}c@{}}Malicious\\ Exploitation\\ on Resource\end{tabular} & EIDOS                                    \\
                               & 2019-10      &  degree $\alpha$    & Fig.~\ref{fig:alpha-mtg-eos}                & Spam                                                                          & defind.io                             \\
                               & 2019-10       & From Account        & Fig.~\ref{fig:percentage-acg-eos}          & Spam                                                                           & defind.io                                         \\
                               & 2019-11       & \# WCC             & Fig.~\ref{fig:cc-cig}                         & Token                                                                             & krownairdrop                                  \\
                               & 2019-05       & \# WCC            & Fig.~\ref{fig:cc-mtg}                         & Spam                                                                          & EOS Global                            \\
                               & 2019-05       & \# SCC            & Fig.~\ref{fig:cc-mtg}                         & Spam                                                                          & EOS Global                                    \\
                               & 2019-05       & \# SCC            & Fig.~\ref{fig:cc-cig}                         & Tool                                                                          & AirDropsDAC                           \\
                               & 2018-11       & $R$               & Fig.~\ref{fig:pearson-mtg}                    & Gambling                                                                      & BET24                         \\
\midrule
\multirow{7}{*}{\textbf{ETH}} & 2016-06       & \# WCC              & Fig.~\ref{fig:cc-mtg}                   & Attack                                                                         & The DAO                                    \\
                               & 2016-10       & \# traces           & Fig.~\ref{fig:traces-mtg}               & \begin{tabular}[c]{@{}c@{}}Malicious\\ Exploitation\\ on Resource\end{tabular} & DoS Attack                                 \\
                               & 2016-10       & From Smart Contract & Fig.~\ref{fig:percentage-mtg-eth}       & \begin{tabular}[c]{@{}c@{}}Malicious\\ Exploitation\\ on Resource\end{tabular} & DoS Attack                                 \\
\bottomrule
\multicolumn{6}{l}{\begin{tabular}[c]{@{}l@{}}* 1LuckyR1fFHEsXYyx5QK4UFzv3PEAepPMK\\ ** 14mUbjiofYY2F6h3ZGUSoTo3kxdqtajVTp\end{tabular}}                                                                                                                                                  
\end{tabular}%
}
\label{table:anomalies-1}
\end{table*}

\newpage
\begin{table*}[h]
\caption{The detected outliers and its corresponding sub-category, as well as the specific reason.}
\centering
\resizebox{0.68\textwidth}{!}{%
\begin{tabular}{lccccc}
\toprule
\textbf{Chain}                 & \textbf{Date} & \textbf{Metric}     & \textbf{Figure}              & \textbf{Sub-category}                                                          & \textbf{DApp / Event}                      \\
\midrule
\multirow{50}{*}{\textbf{ETH}} & 2018-08       & From Smart Contract & Fig.~\ref{fig:percentage-mtg-eth}       & Gambling                                                                       & Last Winner                                \\
                               & 2016-08       &$R$                  & Fig.~\ref{fig:pearson-cig}              & Exchange                                                                       & ReplaySafeSplit                            \\
                               	& 2017-05       &$R$                  & Fig.~\ref{fig:pearson-cig}              & Tool                                                                           & ENS                                        \\
                               & 2018-08       &$R$                  & Fig.~\ref{fig:pearson-cig}              & Gambling                                                                       & Last Winner                                \\
                               & 2019-09       &$R$                  & Fig.~\ref{fig:pearson-cig}              & DeFi                                                                           & NEST Protocol                              \\
                               & 2016-10       & From Smart Contract & Fig.~\ref{fig:percentage-cig-eth}       & \begin{tabular}[c]{@{}c@{}}Malicious\\ Exploitation\\ on Resource\end{tabular} & DoS Attack                                 \\
                               & 2016-10   & \# traces           	& Fig.~\ref{fig:traces} & \begin{tabular}[c]{@{}c@{}}Malicious\\ Exploitation\\ on Resource\end{tabular}	& DoS Attack\\
                               & 2016-10       & \# traces           & Fig.~\ref{fig:traces-cig}               & \begin{tabular}[c]{@{}c@{}}Malicious\\ Exploitation\\ on Resource\end{tabular} & DoS Attack                                 \\
                               & 2019-11       & trace ratio         & Fig.~\ref{fig:traces-cig}               & Game                                                                           & Gods Unchained                             \\
                               & 2018-01       & \# SCC              & Fig.~\ref{fig:cc-cig}                   & Attack                                                                         & Bittrex Hacked                             \\
                               & 2019-02       & \# SCC              & Fig.~\ref{fig:cc-cig}                   & Game                                                                           & CryptoKitties                              \\
                               & 2019-02       & \# SCC              & Fig.~\ref{fig:cc-cig}                   & Token                                                                          & ethairdrop                                 \\
                               & 2019-09       & \# SCC              & Fig.~\ref{fig:cc-cig}                   & DeFi                                                                           & USDT                                       \\
                               & 2018-05       & indegree $\alpha$    & Fig.~\ref{fig:alpha-cig-eth}            & Token                                                                          & EOS                                        \\
                               & 2018-05       & indegree $\alpha$    & Fig.~\ref{fig:alpha-cig-eth}            & Token                                                                          & NePay                                      \\
                               & 2019-06       & degree $\alpha$      & Fig.~\ref{fig:alpha-cig-eth}            & Token                                                                          & MGC TOKEN                                  \\
                               & 2019-06       & degree $\alpha$      & Fig.~\ref{fig:alpha-cig-eth}            & Token                                                                          & VOKEN                                      \\
                               & 2016-10       & From Smart Contract & Fig.~\ref{fig:percentage-acg-eth}       & \begin{tabular}[c]{@{}c@{}}Malicious\\ Exploitation\\ on Resource\end{tabular} & DoS Attack                                 \\
                               & 2016-10       & From EOA            & Fig.~\ref{fig:percentage-acg-eth}       & \begin{tabular}[c]{@{}c@{}}Malicious\\ Exploitation\\ on Resource\end{tabular} & DoS Attack                                 \\
                               & 2018-07       & From EOA            & Fig.~\ref{fig:percentage-acg-eth}       & Spam                                                                           & 0x004b * \\
                               & 2019-05       & From EOA            & Fig.~\ref{fig:percentage-acg-eth}       & Spam                                                                           & 0x8c4b ** \\
                               & 2019-09       & From EOA            & Fig.~\ref{fig:percentage-acg-eth}       & Spam                                                                           & 0x8c4b ** \\
                               & 2016-10       & \# traces           & Fig.~\ref{fig:traces-acg}               & \begin{tabular}[c]{@{}c@{}}Malicious\\ Exploitation\\ on Resource\end{tabular} & DoS Attack                                 \\
                               & 2016-10       & degree $\alpha$      & Fig.~\ref{fig:alpha-acg-eth}            & \begin{tabular}[c]{@{}c@{}}Malicious\\ Exploitation\\ on Resource\end{tabular} & DoS Attack                                 \\
                               & 2016-10       & outdegree $\alpha$   & Fig.~\ref{fig:alpha-acg-eth}            & \begin{tabular}[c]{@{}c@{}}Malicious\\ Exploitation\\ on Resource\end{tabular} & DoS Attack                                 \\
                               & 2016-10       & \# SCC              & Fig.~\ref{fig:cc-acg}                   & \begin{tabular}[c]{@{}c@{}}Malicious\\ Exploitation\\ on Resource\end{tabular} & DoS Attack                                 \\
                               & 2019-10	   & \# WCC	                & Fig.~\ref{fig:cc-mtg}      	& Gambling	& FairWin \\
\bottomrule
\multicolumn{6}{l}{\begin{tabular}[c]{@{}l@{}}* 0x004bd3562a42c8a7394794849b8ff5ad71c527b2\\ ** 0x8c4b7870fc7dff2cb1e854858533ceddaf3eebf4\end{tabular}}  
\end{tabular}%
}
\label{table:anomalies-2}
\end{table*}

\newpage
\begin{table*}[h]
\caption{The detected outliers and its corresponding sub-category, as well as the specific reason.}
\centering
\resizebox{0.68\textwidth}{!}{%
\begin{tabular}{lccccc}
\toprule
\textbf{Chain}                 & \textbf{Date} & \textbf{Metric}     & \textbf{Figure}              & \textbf{Sub-category}                                                          & \textbf{DApp / Event}                      \\
\midrule
\multirow{20}{*}{\textbf{ETH}}  & 2016-11	& \# SCC	& Fig.~\ref{fig:cc-mtg}    	& Exchange	& ReplaySafeSplit\\
                               & 2017-01	& \# WCC	& Fig.~\ref{fig:cc-acg} 	& Exchange	& Yunbi \\
                               & 2017-08  &	\# WCC / \# nodes	 & Fig.~\ref{fig:cc-acg}  &Exchange	& Bittrex \\
                               & 2018-05	& degree $\alpha$   	& Fig.~\ref{fig:alpha-cig-eth} 	& Token	& NePay \\
                               & 2016-05	& degree $\alpha$   	& Fig.~\ref{fig:alpha-mtg-eth}	 	& Exchange & ShapeShift \\
                               & 2016-05	& outdegree $\alpha$		& Fig.~\ref{fig:alpha-mtg-eth} 		 & Token	 & TheDAO\\
                               & 2018-01	& \# traces           	& Fig.~\ref{fig:traces-mtg} 	 & Exchange & Binance \\
                               	& 2018-01	&  trace ratio           	& Fig.~\ref{fig:traces-mtg} 	 & Exchange & Binance \\
                               	& 2018-01	&  trace ratio           	& Fig.~\ref{fig:traces-mtg} 	 & Game &	CryptoKitties \\
                               	& 2018-05  & \# SCC & Fig.~\ref{fig:cc-mtg} 	 & Exchange & Binance \\
                               	& 2018-11	& \# WCC		& Fig.~\ref{fig:cc-mtg}      	 	& Game	& MegaCryptoPolis \\
                               	& 2019-02  &	\# traces           	& Fig.~\ref{fig:traces} 	 & Game &	Dragonereum \\
                               	& 2019-02  &	\# traces           	& Fig.~\ref{fig:traces-cig} 	 & Game &	Dragonereum \\
                               	& 2016-10  & \# SCC           & Fig.~\ref{fig:cc-cig}               & \begin{tabular}[c]{@{}c@{}}Malicious\\ Exploitation\\ on Resource\end{tabular} & DoS Attack \\
                               	& 2016-10  & \# WCC / \# nodes          & Fig.~\ref{fig:cc-acg}               & \begin{tabular}[c]{@{}c@{}}Malicious\\ Exploitation\\ on Resource\end{tabular} & DoS Attack \\
                               	& 2020-03 & 	\# SCC              	 & Fig.~\ref{fig:cc-acg}		& Token &	GasToken.io \\
                               	& 2016-02 & $R$        & Fig.~\ref{fig:pearson-cig}        & Tool  & EthereumAlarmClock \\
                               	& 2016-03 & $R$        & Fig.~\ref{fig:pearson-mtg}        & Exchange  & ShapeShift \\
                               	& 2016-03 & $R$        & Fig.~\ref{fig:pearson-mtg}        & Exchange  & Poloniex \\
\bottomrule
\end{tabular}%
}
\label{table:anomalies-3}
\end{table*}

\newpage
\section{EIDOS \& Spam Advertisements}
\label{sec:appendix:case-study}
In \S\ref{sec:abnormal-behaviors:case-study}, we have briefly introduced the conclusion after we studied the EIDOS project and spam advertisements behavior in EOSIO.
Here, we will provide the full details for them.

\subsection{EIDOS Event}
\label{sec:appendix:case-study:EIDOS}
\begin{figure*}[h]
\centerline{\includegraphics[width=0.7\textwidth]{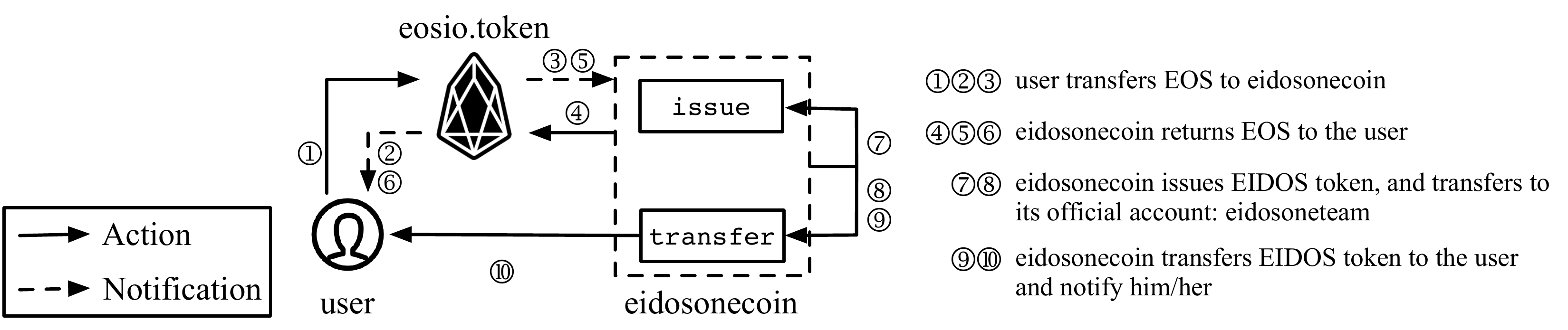}}
\caption{The mechanism of EIDOS.}
\label{fig:eidos-mechanism}
\end{figure*}

As we briefly explained in \S\ref{sec:background:blockchain-evolution}, EIDOS attracts users to transfer EOS to it inexhaustibly. The mechanism of EIDOS is illustrated in Fig.~\ref{fig:eidos-mechanism}.

We can see from Fig.~\ref{fig:eidos-mechanism}, once the user transfers arbitrary amount of EOS by $T_m$ to its main contract: \texttt{eidosonecoin} (see steps 1 to 3), the EOS would be immediately returned by another $T_m$ (see steps 4 to 6). Then, EIDOS would initiate a $T_c$ that invokes the \texttt{transfer} function in its owned smart contract (see steps 9 and 10). In the \texttt{transfer} function, EIDOS would increase the user's balance of EIDOS token whose value is not related to the amount of EOS in the previous $T_m$. Users can spend these EIDOS tokens in exchanges or other contracts that accept EIDOS token.
Note that the EIDOS token is issued by the \texttt{issue} function and part of them will be transferred to its official team: \texttt{eidosoneteam} (see steps 7 and 8). Therefore, there also exists an another two $T_c$s which invoke \texttt{issue} and \texttt{transfer}. 
However, the issuance of EIDOS is not executed each time, thus the ratio of $T_m$ to $T_c$ is roughly two to one, which were illustrated by the `trace ratio' in Fig.~\ref{fig:traces-mtg} and Fig.~\ref{fig:traces-cig}.

\begin{figure}[h]
     \centering
     \begin{subfigure}[t]{0.49\columnwidth}
         \centering
         \includegraphics[width=\textwidth]{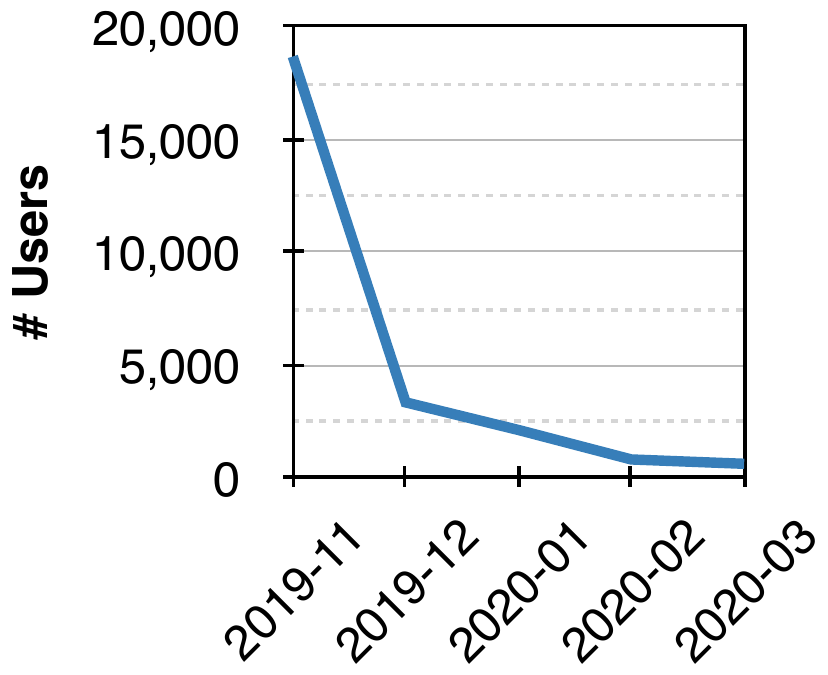}
         \caption{Amount of users.}
         \label{fig:eidos-user}
     \end{subfigure}
     \begin{subfigure}[t]{0.49\columnwidth}
         \centering
         \includegraphics[width=\textwidth]{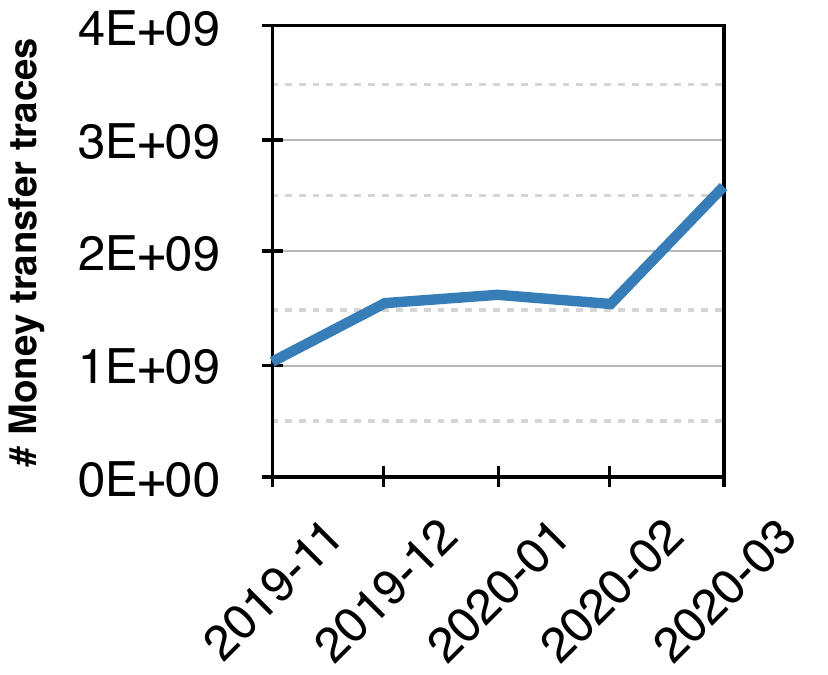}
         \caption{Amount of $T_m$.}
         \label{fig:eidos-trace}
     \end{subfigure}
     \begin{subfigure}[t]{0.49\columnwidth}
         \centering
         \includegraphics[width=\textwidth]{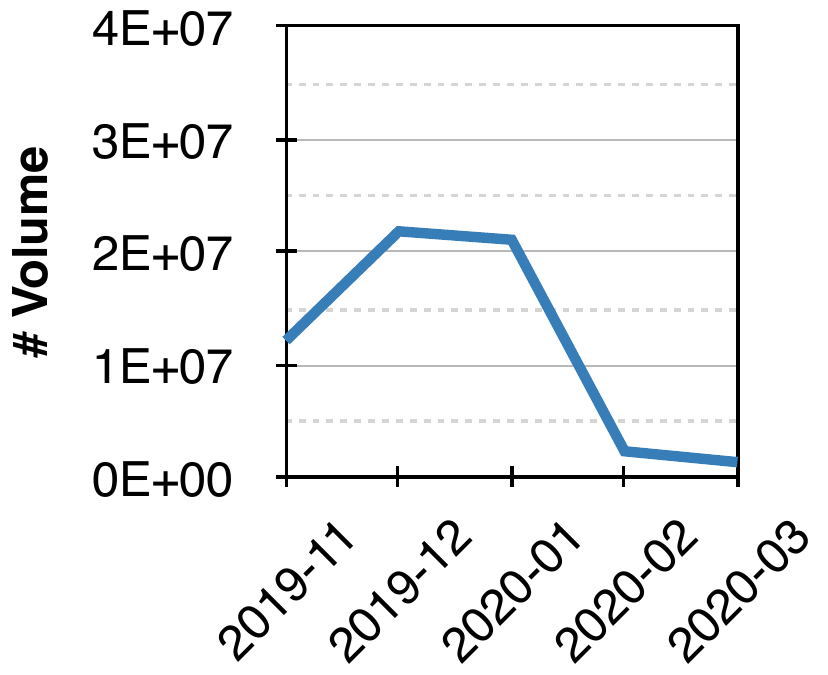}
         \caption{Transferred volume in EOS.}
         \label{fig:eidos-volume}
     \end{subfigure}
     \begin{subfigure}[t]{0.49\columnwidth}
         \centering
         \includegraphics[width=\textwidth]{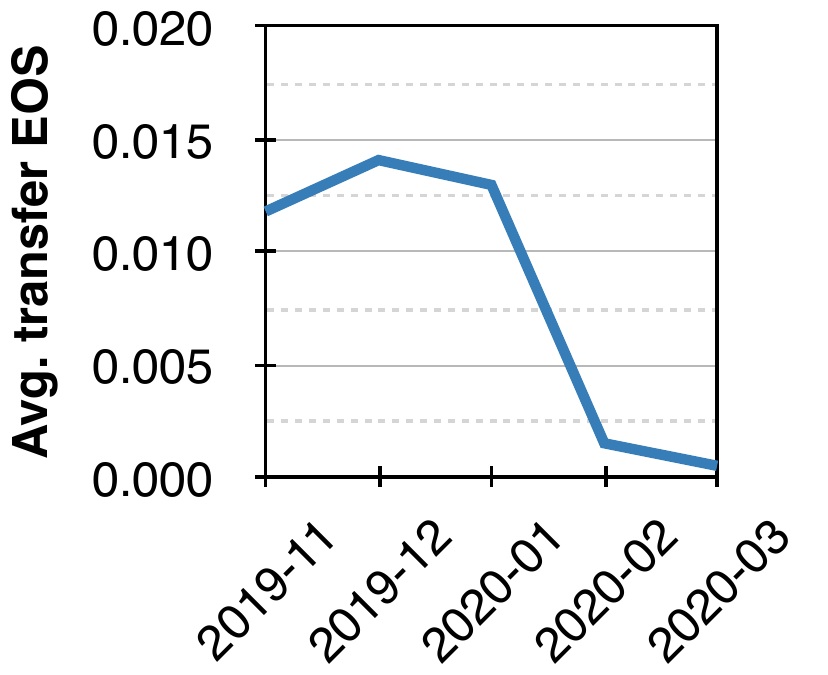}
         \caption{Average transferred EOS per $T_m$.}
         \label{fig:eidos-eos-per-trace}
     \end{subfigure}
        \caption{The statistics of EIDOS project.}
        \label{fig:eidos-statistics}
\end{figure}

According to our statistics, in November 2019, the number of traces in EOSIO has jumped by 1009.82\% over the preceding month, reaching up to 3.21 billion pieces. Therefore, we further extract all the traces related to EIDOS, and calculate several related metrics as shown in Fig.~\ref{fig:eidos-statistics}.

We see that the number of users decreased dramatically in Mar. 2020: the number has dropped to 643 (around 95.13\% compared to Nov. 2019). In the meanwhile, the number of related $T_m$ has increased by 150.43\%. These two opposite changes indicate that the players of EIDOS tend to be specialized. In other words, some accounts started to a provide service, which allows users to invest their owned resources and get dividends back. Actually, there indeed existed some service providers~\cite{eidos-miner-1, eidos-miner-2}. These can also be observed from the $\alpha$ of outdegree distribution during the corresponding time period (see Fig.~\ref{fig:alpha-mtg-eos}).

Moreover, these players were becoming \textit{rational}. As we have explained in \S\ref{sec:abnormal-behaviors:case-study:EIDOS}, the amount of the returned EIDOS token has nothing to do with the amount of invested EOS. Thus, as we can see from Fig.~\ref{fig:eidos-volume} and Fig.~\ref{fig:eidos-eos-per-trace}, after an ephemeral increase in the volume of invested EOS and the amount of EOS per invested traces, the two curves start to go down sharply. In Mar. 2020, the average EOS per trace was as low as $6 \times 10^{-4}$, while $1 \times 10^{-4}$ is the minimum allowed transferred amount in EOSIO and the most rational choice.
Actually, according to our data, in Nov. 2019, 99.80\% of $T_m$ were valued below 1 EOS. But in Oct. 2019, this percentage was up to 74.81\%.
Therefore, we conclude that EIDOS has captured the whole EOSIO ecosystem via meaningless $T_m$ and $T_c$ transactions, thereby slowing the TPS of other users.

\subsection{Spam Advertisement}
\label{sec:appendix:case-study:spam}
During the analysis of Fig.~\ref{fig:alpha-mtg-eos}, we have observed two strange spikes, located in Mar. 2019 and Oct. 2019, in the $\alpha$ of outdegree distribution. After the analysis of our method mentioned in \S\ref{sec:abnormal-behaviors:method} and manual verification, we figured out these two outliers resulted from \textit{spam advertisements}.

To be specific, taking the advantage of memo field in EOSIO transactions~\cite{eos-memo} where the initiator can freely write anything, an account or a smart contract is able to initiate lots of $T_m$ carrying with $1 \times 10^{-4}$ EOS and bait-and-switch advertisement to cover as many users as possible.

Based on our investigation, we proposed a detection methodology. Specifically, our methodology traverses the $MTG$ and $CIG$ for each month, and detects if node $N$ obeys to the following rules:
\begin{enumerate}
	\item The average amount of transferred EOS to each user initiated from $N$ is no more than $x$;
	\item $N$ invokes up to $y$ pieces of $T_m$ to each user within a single month;
	\item More than $z$ users who received $T_m$ from $N$ did not in turn initiates $T_m$ or $T_c$ to $N$;
	\item The memo should carry the leading content, typically is an exaggerated statement and a URL.
\end{enumerate}

\begin{figure*}[h]
\centerline{\includegraphics[width=0.7\textwidth]{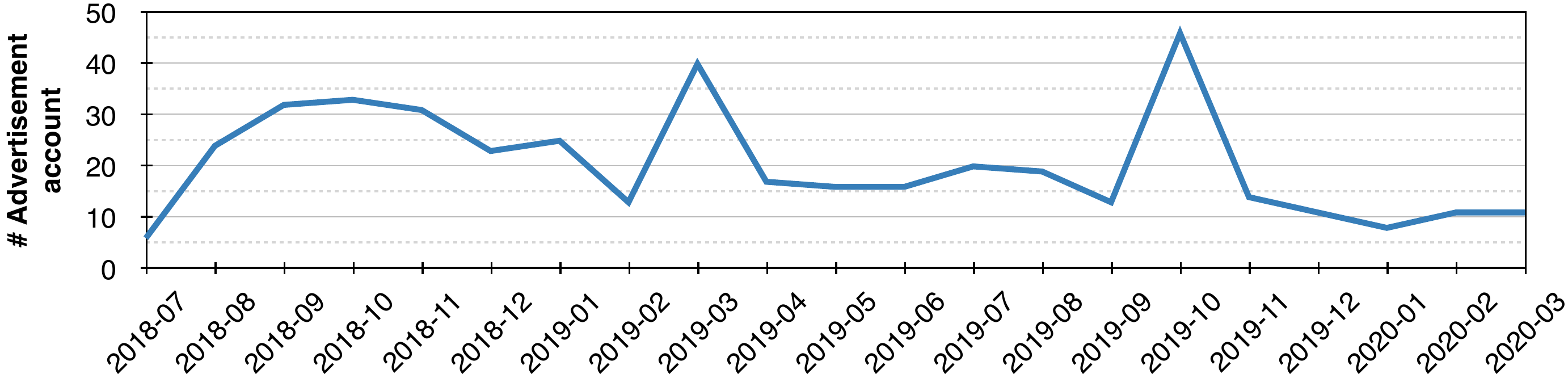}}
\caption{The distribution of these 206 detected advertisement accounts according their first appearance time.}
\label{fig:ad-distribution}
\end{figure*}

The $x$, $y$ and $z$ are heuristically set as 0.001, 30, and 500, which are relative conservative and may lead to false positives.
Thus we further conduct a manual verification focusing on the memo content and the patterns of behavior.
We discover \textbf{206} distinct accounts in total that have initiated spam advertisements at least for one month. The distribution of the first appearance time for these 206 accounts is shown in Fig.~\ref{fig:ad-distribution}.
We can see the obvious two spikes correspond to the outliers located in Mar. 2019 and Oct. 2019, respectively.
However, except for these two spikes and the peak in the late-2018 (caused by gambling DApps), the advertisement accounts are not so active, especially after EIDOS.

\begin{figure}[h]
\centerline{\includegraphics[width=\columnwidth]{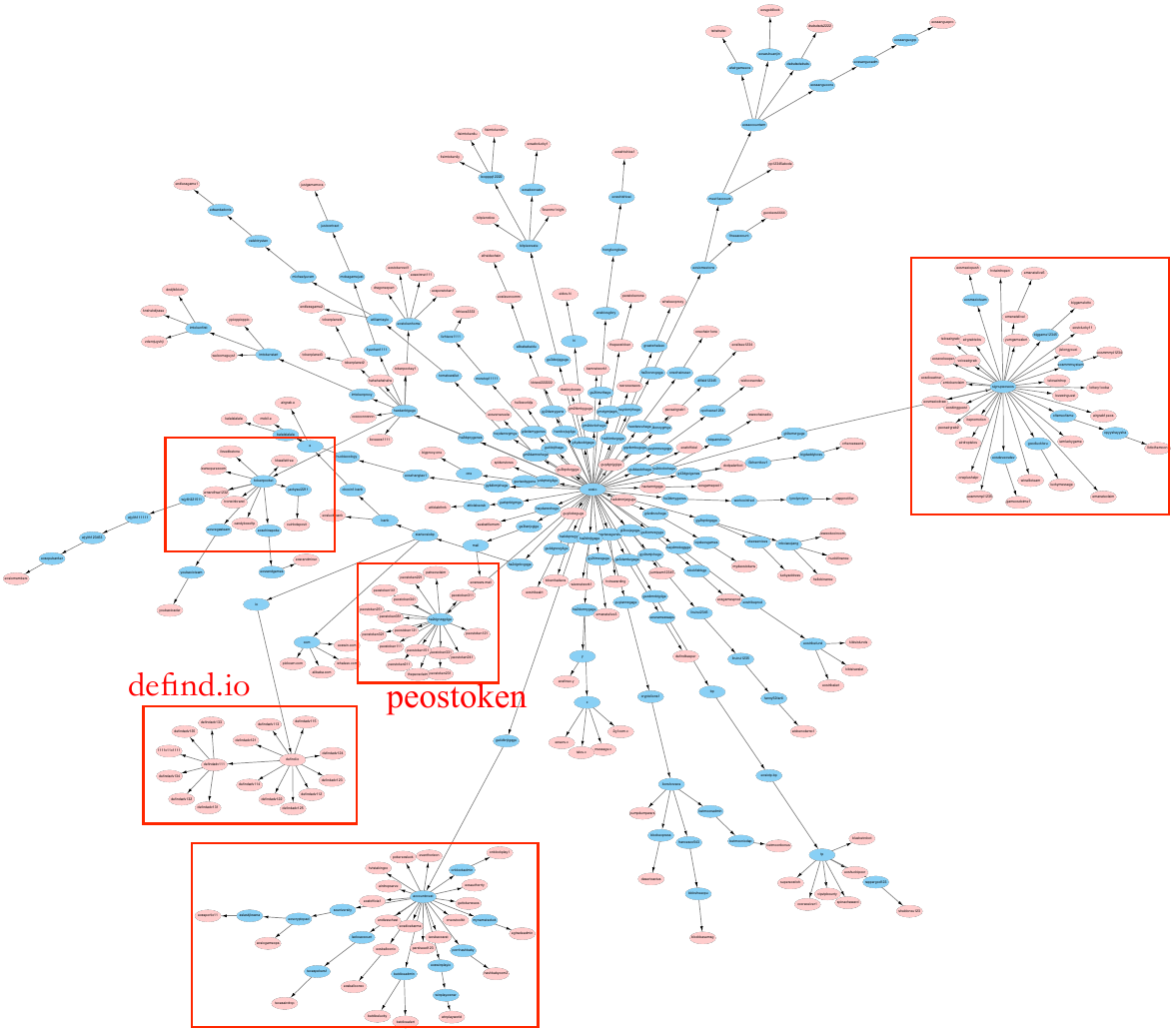}}
\caption{The family tree of spam advertisement related accounts, where the red nodes are advertisement nodes.}
\label{fig:ad-family-tree}
\end{figure}

After investigation, we figure out these two spikes are caused by two clusters of advertisement accounts: \texttt{peostoken} family and \texttt{defind.io} family.
Therefore, we have built a family tree to explore the creation relationship, taking advantage of our $ACG$, which is shown in Fig.~\ref{fig:ad-family-tree}.
We observe that there are clusterings (framed by red boxes in the figure) in which the proportion of advertisements nodes are relatively high. We have marked the \texttt{peostoken} family and \texttt{defind.io} family in Fig.~\ref{fig:ad-family-tree}, which were the reasons behind the two outliers. These two clusters are composed of accounts with multi-level creation relationships.

In addition, we have observed some interesting characteristics for spam advertisement accounts:
\begin{itemize}
	\item The language in memo content is mainly in English and Chinese, as well as a small percentage of Korean and Japanese.
	\item It is possible for the same account to send out memos with different content under different time periods, and even change the language used in the memo.
	\item Most of the advertisements are promoting their own gambling games, and some are promoting their own tokens or services (e.g., airdrop services). Moreover, advertisements promoting DeFi-related DApps have been appearing since Sep. 2019.
	\item \texttt{eostribalert} and \texttt{eostribealrt} are a pair of accounts that launch advertisements. The memo in the latter one contains a URL and convinces users to apply for TLOS token. Immediately after the former one also starts sending out a large number of $T_m$ containing a memo stating that the URL is fraudulent information. However, the difference of number of $T_m$ they sent is very large, 5.3K and 101K times, respectively.
\end{itemize}

We conclude that spam advertisement behavior was popular in EOSIO. They mainly target users who speak English and Chinese. Moreover, those advertisement accounts usually have relationships in terms of account creation behavior.

\end{document}